\documentclass[iop]{emulateapj}

\usepackage{pdfpages}

\newcommand{\fe}{[Fe/H]}
\newcommand{\kms} {km s$^{-1}$}

\newcommand{\lgg}{log g}
\newcommand{\ug}{{\it ugriz}~}
\newcommand{\ugp}{$u'g'r'i'z'$~}

\newcommand{\ebv}{$E (B-V)~$}
\newcommand{\tef}{T$_{\rm eff}~$}
\newcommand{\gr}{$ (g-r)_0~$}

\shortauthors{Morrison et al.}

\begin{document}

\title{Globular and Open Clusters Observed by SDSS/SEGUE: the Giant Stars}

\author{Heather L. Morrison\altaffilmark{1}, 
Zhibo Ma\altaffilmark{1}, 
James L. Clem\altaffilmark{2}, 
Deokkeun An\altaffilmark{3}, 
Thomas Connor\altaffilmark{1}$^,$\altaffilmark{4},
Andrew Schechtman-Rook\altaffilmark{1}$^,$\altaffilmark{5}, 
Paul Harding\altaffilmark{1},
Luca Casagrande\altaffilmark{6},
Constance Rockosi\altaffilmark{7},
Brian Yanny\altaffilmark{8},
Timothy C. Beers\altaffilmark{9}$^,$\altaffilmark{10},
Jennifer~A.~Johnson\altaffilmark{11},
Donald P. Schneider\altaffilmark{12}$^,$\altaffilmark{13}
}

\affil{$^1$ Department of Astronomy, Case Western Reserve University,
  Cleveland, OH 44106}
\affil{$^2$ Department of Physics, Grove City College, 100 Campus Dr., Grove City, PA  16127}
\affil{$^3$ Department of Science Education, Ewha Womans University, Seoul 120-750, Republic of Korea}
\affil{$^4$ Department of Physics and Astronomy, Michigan State University, 567 Wilson Rd, East Lansing, MI 48824}
\affil{$^5$ Department of Astronomy, University of Wisconsin, 475 North Charter Street, Madison, WI 53706}
\affil{$^6$ Research School of Astronomy and Astrophysics, Mount Stromlo Observatory, The Australian National University, ACT 2611, Australia}
\affil{$^7$ UCO/Lick Observatory, University of California, Santa Cruz, 1156 High St., Santa Cruz, CA 95064, USA.}
\affil{$^8$ Fermi National Accelerator Laboratory, PO Box 500, Batavia IL 60510, USA}
\affil{$^9$ Department of Physics, University of Notre Dame, 225 Nieuwland Science Hall, Notre Dame, IN 46656, USA}
\affil{$^{10}$ JINA: Joint Institute for Nuclear Astrophysics, University of Notre Dame, Notre Dame, IN 46556}
\affil{$^{11}$ Department of Astronomy, 
Ohio State University, 140 West 18th Avenue, Columbus, OH 43210, USA.}
\affil{$^{12}$ Department of Astronomy and Astrophysics, The Pennsylvania State University,  University Park, PA 16802}
\affil{$^{13}$ Institute for Gravitation and the Cosmos, The Pennsylvania State University, University Park, PA 16802}

\begin{abstract}

We present $griz$ observations for the clusters M92, M13 and NGC 6791
and $gr$ photometry for M71, Be 29 and NGC 7789. In addition we
present new membership identifications for all these clusters, which
have been observed spectroscopically as calibrators for the SDSS/SEGUE
survey; this paper focuses in particular on the red giant branch stars
in the clusters. In a number of cases, these giants were too bright to
be observed in the normal SDSS survey operations, and we describe the
procedure used to obtain spectra for these stars.  For M71, also
present a new variable reddening map and a new fiducial for the $gr$
giant branch. For NGC 7789, we derived a transformation from \tef\ to
$g-r$ for giants of near solar abundance, using IRFM \tef\ measures of
stars with good \ug\ and 2MASS photometry and SEGUE spectra. The
result of our analysis is a robust list of known cluster members with
correctly dereddened and (if needed) transformed $gr$ photometry for crucial
calibration efforts for SDSS and SEGUE.

\end{abstract}


\section{Introduction}

Calibrations which relate observables such as stellar photometry and
spectroscopy to fundamental stellar parameters are a vital part of any
survey.  The Sloan Digital Sky Survey \citep[SDSS:][]{york00} provided
imaging in five passbands (\ug) for 14,555 square degrees of the sky,
using a dedicated imager \citep{gunn98} on the SDSS 2.5m telescope
\citep{gunn06}. This photometric database was complemented by
spectroscopic observations using a multi-object spectrograph
\citep{smee13}. The original purpose of the SDSS survey was to map the
extragalactic universe by obtaining spectra of one million galaxies
and one hundred thousand quasars. However, because of a number of
important, serendipitous discoveries on the Milky Way, two surveys
(SEGUE-1 and -2) which focused on the stellar populations in the Milky
Way were carried out as extensions to the original SDSS. SEGUE-1
\citep{yanny09} acquired data from 2005 through 2008, and SEGUE-2
\citep[][Rockosi et al., in preparation]{eisenstein11} in 2008 and
2009.

Because the SDSS $ugriz$ photometric system \citep{fukugita} was
originally designed for the study of galaxies and quasars rather than
stars (focusing on avoiding strong sky lines rather than on features
in a zero redshift stellar spectrum), one of the important tasks for
the SEGUE survey was to obtain observations in $ugriz$ for
well-studied objects with known stellar parameters. This program has
allowed us to understand how \tef, \lgg\ and [Fe/H] map into the SDSS
colors. Such understanding is particularly important when studying
stars from minority populations in the Galaxy such as its halo, as
there are often [Fe/H] and luminosity terms in transformations from
other photometric systems to $ugriz$, particularly in the $u$ and $g$
filters \citep[for example, see Figure 10 of][]{yanny09}. Open and
globular clusters are particularly suitable as calibrators because
they provide many objects with the same values of [Fe/H].

The first order of business in using star clusters is to obtain a good
color-magnitude diagram (CMD) in the appropriate filter set. Because
the SDSS photometric pipeline \citep{EDR} does not perform well in
crowded fields, \citet{an08} performed DAOPHOT photometry of open and
globular clusters imaged by SDSS. \citet{an08} then provided accurate
fiducial sequences for 17 globular and 3 open clusters covering a
metallicity range from \fe=--2.4 to +0.4. However, since the SDSS
camera saturates at around $g=14.5$, and a number of these clusters
have giant branches reaching significantly brighter than this limit,
it was necessary to use observations from other telescopes.  The \ugp
system is defined by the same filters as the \ug system and was
intended to simplify observations in \ug from other telescopes
\citep{smith02,tucker}. However, because the filters are in vacuum in
the SDSS imager but not when used in other telescopes, the
two photometric systems are in fact different. We used the \ugp observations
of the bright giant branches of these clusters \citep{clem08}, transforming
them to $ugriz$ using the transformations of \citet{tucker}.

The next step in the use of clusters as calibrators is to make sure
that the stars we study are in fact cluster members. Spectroscopic
observations provide velocities and other useful discriminants of
membership, and the SEGUE survey obtained spectra of stars in 13
clusters \citep{lee08b,smolinski}.  These two papers used velocity,
position on the CMD and the metallicity of the star as measured by the
SEGUE Stellar Parameters Pipeline (SSPP:
\citet{lee08a,lee08b,carlos08,smolinski}) as membership criteria.

Our particular interest is the cluster red giant branches. For
technical reasons described below, the giant branch stars were not
well covered in the previous tests of the SSPP by \citet{lee08b} and
\citet{smolinski}.  The next paper in this series (Morrison et al
2015, in preparation) shows how we used the clusters described in this
paper to test the values of [Fe/H] and log g for cluster members, and
adds an additional luminosity discriminant (the Mg index) to enhance
the SSPP's ability to identify red giant stars. A number of the
clusters described here either have low radial velocities or are
located at low galactic latitude, making it difficult to distinguish
cluster members from foreground disk stars using only velocity. We
have chosen to identify cluster members using a different set of
criteria than \citet{lee08b} and \citet{smolinski}: while we both use
the SEGUE velocities and the position on the cluster CMD, we have
chosen not to use the SSPP metallicity, and have added another
powerful discriminant: the stellar proper motion. Proper motion data
are available for all but one clusters we study. In one particularly
recalcitrant case, Berkeley 29, where proper motions were not
available, we used the velocity and CMD criteria, then rejected
foreground dwarfs by visual inspection of the spectra.

M71 is a particularly important and difficult case. It is the only
well-studied, nearby cluster with an intermediate metallicity which is
accessible from the Northern Hemisphere. Unfortunately, it also has
variable reddening across its face. Thus we needed to derive
individual reddening estimates for different regions of the cluster
\citep[using the photometry of][]{clem08} before we could produce a
cluster CMD suitable for producing a fiducial for the giant branch.

In this paper we present, for each cluster, a CMD showing the stars
which are likely members and were observed spectroscopically by SEGUE,
and a table of accurate coordinates and other information on these
likely cluster members.  We also describe the SEGUE observations of
cluster stars when the cluster focal plane fiber plug-plates (plates
hereafter) differed in observational procedure from the usual survey
plates. In addition we provide a table of our derived spatial
reddening offsets for M71 and a transformation between \tef\ and $g-r$ for
near solar abundance giants.

\section{Clusters used for Calibration}

The SEGUE project observed a number of globular and open clusters for
calibration purposes. For calibration of the red giants, we selected
 the globular clusters M92, M13 and M71 (spanning metallicities
from --2.4 to --0.8) and the open clusters Be 29, NGC 7789 and NGC 6791,
whose [Fe/H] values range from --0.4 to +0.4. In all but one case, the
clusters are within the SDSS footprint and so $ugriz$ photometry is
available for the cluster stars.

The SDSS cluster images were analyzed using DAOPHOT \citep{daophot} by
\citet{an08} because the SDSS photometric pipeline was not designed to
handle crowded fields.  However, in most cases the cluster giant
branch stars were too bright to be observed in the standard $ugriz$
system (defined by the data taken on the SDSS 2.5m) because they are
saturated in the SDSS exposures. In these cases we used the \ugp
photometry described in \citet{clem08}, transforming using the
equations of \citet{tucker}. \citet{an08} checked these
transformations using data available in both \ug and \ugp and found
agreement at better than the 2\% level for all clusters except NGC
6791, where stars at the tip of the giant branch were redder in
standard \ug than in transformed \ugp by 0.05 to 0.10 magnitudes.
This discrepancy becomes particularly significant redder than
$g-r$=1.0.  Fortunately, NGC 6791 is sufficiently distant that we are
able to use the SDSS \ug for all its red giants.

We summarize the values of distance modulus, \fe~ and \ebv that we adopted for
these clusters, along with the sources of these measurements, in Table
\ref{clusterdata}. For the globular clusters we use the metallicity scale
of \citet{kraftivans03}, based on FeII lines.

In addition, we checked the list of globular cluster variables compiled by Christine Clement\footnote{found at
http://www.astro.utoronto.ca/~cclement/read.html} to see if any of the
stars that we observed were long-period variables, since their use as
calibrators would be unwise. We found that there were no known LPVs in
our sample of globular cluster stars.

The SEGUE-1 survey is described in \citet{yanny09}. The survey obtained low-resolution (R$\sim$1800) spectra for the wavelength region from 3800 to 9000 \AA. 
Each spectroscopic pointing had a bright and faint plug-plate (plate hereafter),
with exposure times of typically one and two hours respectively. This procedure
allowed us to reduce the effect of scattered light from bright stars
in adjacent fibers to fainter stars by limiting the magnitude range on
a given plate. For the clusters, we used a more flexible setup to
attempt to obtain spectra with good S/N for as many stars as possible.

 For many clusters, more than one plate was designed and
 observed. Table \ref{clusterplates} summarizes the information on the
 plates taken for each cluster discussed in this paper; often a
 cluster had both a “bright” and a “faint” plate designed. For the
 brightest stars in M92 and M13, only a very short exposure, of
 duration 1-–2 minutes, was needed. This caused a possible problem
 with our spectroscopic reduction pipeline, since it uses night sky
 lines in the spectra to check the wavelength calibration, and such
 short exposures are too short to properly expose the sky lines. We
 evolved the following procedure in order to make such observations
 process correctly in the pipeline. The brightest stars were observed
 by drilling their fiber holes on the bright plate at a position
 offset by 0.02 degrees in RA (cos(Dec))$^{-1}$. This offset is
 $\sim$1.5 arcmin for M92 and M13. While the rest of the stars on the
 bright plate were observed, a sky spectrum accumulated in these
 fibers. When the “bright” exposure was finished, the telescope was
 moved by this offset, taking the regular stars away from their fibers
 and placing the brightest stars on the fibers which had been
 accumulating sky photons. The plate was then exposed for an
 additional short time.  The coordinates in the SDSS database have
 been corrected for these offsets.
The referee has asked whether starlight could have accumulated in these
offset fibers during the 'sky' exposures. We note that we avoided placing
fibers on the bright central regions of both clusters, with the nearest star
more than 2' from the cluster center in M92 and $\sim$1' from the less
crowded center of M13. Thus it is unlikely that any light from cluster stars will
contaminate the M92 offset fibers, and a little more likely for M13, where
we note that in some of the stars observed with offsets, the signal-to-noise ratio is
 lower than expected at the blue  end of the spectrum (4000\AA\ and below).

In addition, this offset procedure may introduce additional
uncertainties on the radial velocities.  \citet{yanny09} quantify the
radial velocity uncertainty for SEGUE survey plates as a function of
both $g$ magnitude and S/N. The CMDs of the clusters in the following
section show that SEGUE observed stars as faint as $g$=20, with a wide
range of $g-r$ color. At magnitudes greater than $g$=19, velocity
errors can be 20 \kms\ or more. However, the red giants in the
clusters, the major object of this paper, are redder and in general
much brighter. For all clusters except Be 29, the giants have $g<16$,
which give radial velocity errors around 5 \kms\ for the red
giants. Be 29's giants can be as faint as 19, and so we expect larger
velocity errors for this cluster of up to 10 \kms. The two clusters
with offset observations (M92 and M13) may have an additional error of a few \kms\ introduced for the brightest stars.

The spectroscopic observations of the near solar metallicity open
cluster NGC 7789 were targeted on the giant branch only, 
because there are already good SEGUE spectroscopic observations of
stars below the main-sequence turnoff in two open clusters with [Fe/H]
close to solar: M67 and NGC 2420.

\section{Clusters with Proper Motion Data}

We used the proper motions of Cudworth and collaborators for M92, M13,
M71 \citep[][respectively]{rees92,cudmonet79,cud85} and for NGC 6791 (Cudworth,
private communication). 
For NGC 7789 we used the proper motions of \citet{mcnamara81}.

Because the radial velocity zeropoint was uncertain for some of the
brighter cluster plates, we first examined the proper motion members
(those with membership probability greater than 70\%) to obtain a clean
radial velocity distribution for the cluster. We then used this to
find the optimal range of radial velocities for cluster membership selection. 

\subsection{M92}

\begin{figure}[h!]
\centering
\includegraphics[scale=0.45]{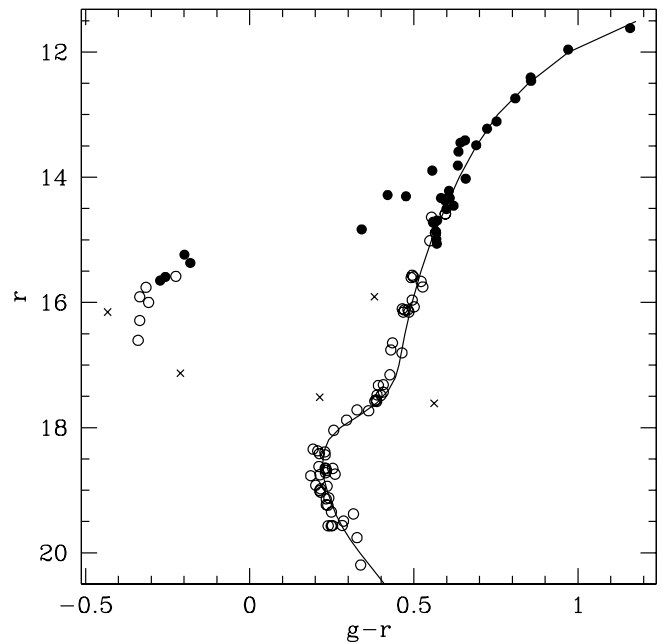}
\caption{$gr$ CMD of the globular cluster
  M92, using data from \citet{an08} and \citet{clem08}.  All points
  plotted are radial velocity members. Stars with proper motion
  membership probabilities higher than 70\% are shown as filled
  circles, while stars with no proper motions available are shown with
  open circles. Crosses are stars which are classified as non-members
  because of their position in the CMD. The solid line is the fiducial
  of \protect \citet{clem08}, transformed to $gr$ using the transformation of
  \protect \citet{tucker}.}
\label{m92}
\end{figure}

For the most metal-poor globular cluster in our dataset, M92, the
giant branch tip is at $r \sim 11.5$, so most of the red giant
branch is saturated in the SDSS photometry. We therefore used both the
 $ugriz$ photometry of \citet{an08} (reference run 4682 plus run 5237, transformed to the reference frame using the transformations given in \citet{an08}) and the transformed
photometry of \citet{clem08} to construct the color-magnitude diagram (Figure \ref{m92}).

SDSS photometric reductions have improved over the years of the
survey. An important advance occurred between Data Releases (DR) 7
and 8: what is known as the ``Ubercalibration'' \citep{nikhil}.  This
technique solves for the photometric calibration parameters using all overlapping
observations.  When \citet{an08} first made the DAOPHOT reductions of
the cluster data available, they calibrated these reductions to
DR7. Subsequently, \citet{an13} calculated the offsets to apply in
order to put the cluster photometry on the DR8 system, and we have
applied these offsets to the An et al.\ photometry given in Table \ref{m92mem}.

 Proper motions from \citet{rees92}  are available for all giants above the level of the horizontal branch.
We chose a radial velocity range of --123 to --103
\kms\ for cluster members. Coordinates, $griz$ photometry and its source, SEGUE radial velocities and proper motion membership probabilities for each member are given in Table \ref{m92mem}. 

\subsection{M13}

\begin{figure}[h!]
\centering
\includegraphics[scale=0.45]{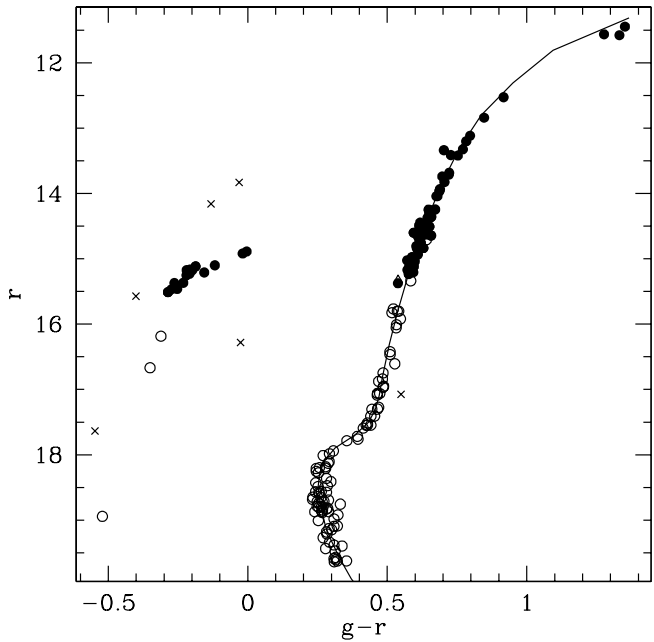}
\caption{$gr$ CMD of the globular cluster M13, using data from
  \citet{an08} and \citet{clem08}.  All points plotted are radial
  velocity members. Symbols have the same meaning as in Figure
  \ref{m92}. The solid line shows the transformed fiducial of
  \protect \citet{clem08}. }
\label{m13}
\end{figure}

M13 also has  bright giants, so the color-magnitude diagram
shown in Figure \ref{m13} uses photometry from both \citet{an08} and
\citet{clem08}. For the photometry from \citet{an08}, we used runs 3225 and 3226, correcting run 3226 to the reference run (3225) using the corrections given in \citet{an08}, 
and then applied the ``Ubercalibration'' corrections given in Table 1 of \citet{an13}.
We chose a radial velocity range of --251 to --239 \kms\ to
select radial velocity members for this cluster. Proper motions are 
available for all of the stars on the giant branch above the horizontal branch.

Data on individual cluster members are given in Table \ref{m13tab}.

\subsection{M71}

Unlike the two globular clusters previously discussed, M71 is a disk
globular cluster in a low-latitude field with variable reddening
\citep[see, e\.g\.][]{luca10}. However, it is one of the few clusters in this
metallicity range accessible from the North. Its low galactic latitude makes
membership decisions more difficult because of the large number of
foreground disk stars. In addition, M71's radial velocity is closer to that
of the field stars because of its disk-like orbit: the difference is only
$\sim -20$ km s$^{-1}$, compared to the M92 and M13 radial velocities which
are 100 \kms\ (or more) different from the field star radial velocities.

Comparison of the giant branch fiducial of \citet{clem08} for M71 with
other cluster fiducials suggests that the shape of the \citet{clem08}
fiducial is slightly incorrect, presumably because of the larger
probability of field star contamination on M71's giant branch, and
because even genuine members have variable reddening and thus they do not
trace a tight fiducial.
We have therefore constructed a new fiducial for the M71 giant branch
in $g$ and $r$, using only stars which are likely members, and
have tightened up the CMD by estimating the variable reddening across
M71's field, using the position of the main sequence turnoff in Clem's
accurate photometry.

To construct the CMD for M71 using likely
members only, we started with stars which had more than a 70\%
probability of membership from the proper motions of \citet{cud85} and
from unpublished data kindly made available by Kyle Cudworth for the
fainter stars. These data reach more than a magnitude below the
horizontal branch, so are ideal for our purposes. We use the
photometry of \citet{clem08}, converted to $gr$ using the
transformations in \citet{tucker}.

For radial velocity membership data, in light of the small
difference between M71's velocity and that of contaminating field
stars, a more accurate velocity catalog was extremely helpful: Tad
Pryor (private communication) kindly provided unpublished velocity data for almost all stars
on or above the horizontal branch in M71. These data were obtained
with high-resolution spectrographs on the DAO 48-inch and the KPNO 4m
and have errors $\leq$ 1 \kms. There were also multiple
observations for many of the stars, allowing likely binaries to be
flagged via their radial velocity variability.  The higher velocity accuracy
allowed us to use a narrower window to define velocity
membership: --20 to --27 \kms. We also rejected one star with both
radial velocity and proper motion suggesting membership (star 1-1, on
plate/MJD/fiber 2333/53682/165) but with radial velocity variations which
may be due to binarity.

To estimate variable reddening values across the field of M71, we used
the M71 photometry of \citet{clem08} to map the position of the
main sequence turnoff across the field, using  $g-i$  to provide
a more sensitive measurement.  We divided the region near M71 into
square regions of size 50 arcsec, plotted the CMD near the turnoff for
each region, and then overlaid the \citet{clem08} M71 fiducial,
varying the reddening offset by eye until we obtained the best
fit. The scatter around the fitted fiducial gives an estimate of the
remaining variation in reddening inside the 50 arcmin square field
(because the Clem et al photometry was internally accurate to
significantly better than 0.01 mag.\ at these bright magnitudes). This
scatter had a range of 0.020 to 0.075 in $g-i$ around the fiducial
(equivalent to up to 0.05 in $E(g-r)$).  Our reddening offsets are
given in Table \ref{redofftab} and vary between +0.07 and --0.03 in
E(B--V) over the M71 field, which is 9.2 arcmin on a side.
 
\begin{figure}[h!]
\centering
\includegraphics[scale=0.45]{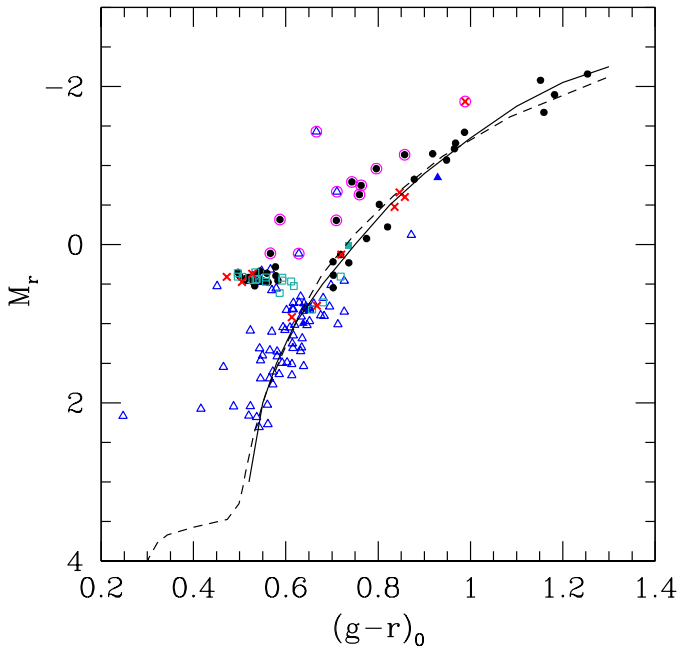}
\caption{CMD of the globular cluster M71, using data from
  \citet{clem08} transformed to $gr$ using the transformations of
  \citet{tucker}. Only stars with more than a 70\% probability of
  proper motion membership are plotted. Solid symbols denote stars
  where variable reddenings have been estimated and applied, while
  open symbols use $E(B-V)$=0.28. Black circles show stars with
  multiple radial velocity observations that are radial velocity
  members without variable velocities. Red crosses show stars with
  radial velocity variability suggesting that the star may be part of
  a binary system or a long-period variable. Blue/green squares
  indicate stars with one radial velocity measurement which are
  velocity members. Likely AGB stars are marked with a large magenta
  circle. Blue triangles denote stars without radial velocity
  measurements. The solid black line traces our new M71 giant branch
  fiducial, and the dotted line the \protect \citet{clem08} fiducial. }
\label{m71cmdfid}
\end{figure}

Applying these reddening offsets to the confirmed cluster members (and
using the global reddening and distance modulus values given in Table
\ref{clusterdata}) results in the CMD shown in Figure \ref{m71cmdfid},
which can be compared with Figure 12 of \citet{clem08}. (We used
Clem's photometry in this case because the stars on M71's giant branch
were either saturated or close to saturated in the original SDSS
exposure.) Our CMD is significantly cleaner, with foreground stars
removed and the red giant branch, red horizontal branch (RHB) and
asymptotic giant branch more clearly visible.  Likely AGB stars are
circled in magenta in the Figure: there are 11 of them. We show our
improved fiducial for the M71 giant branch in the Figure. This
fiducial is tabulated in \citet{xue14}.

\begin{figure}[h!]
\centering
\includegraphics[scale=0.45]{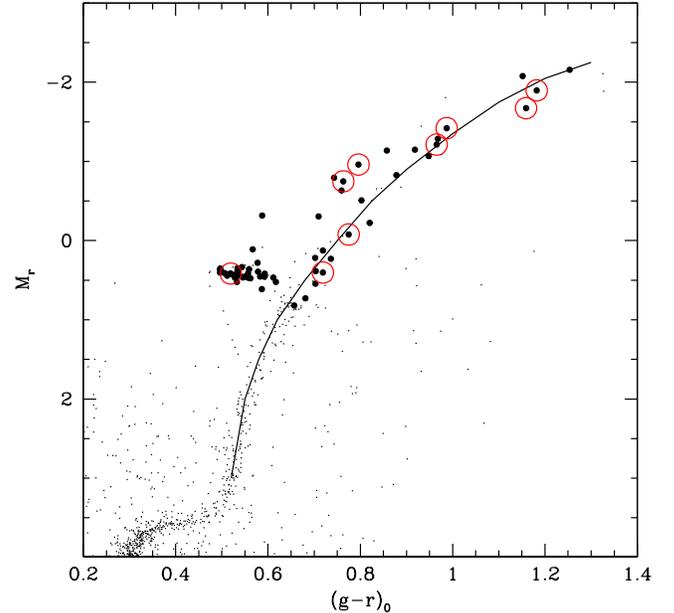}
\caption{CMD of the globular cluster M71, using data from
  \citet{clem08}. Stars without velocity data are shown as small
  black dots, while stars with proper motions and radial velocities
  indicating membership are shown as large black dots.  Stars observed
  by SEGUE are shown by large red circles. }
\label{m71}
\end{figure}

Table \ref{m71tab} lists our cluster members for M71. We also list the
unpublished velocities of Pryor and collaborators, our adopted reddening
offsets in E(B-V) for each of these stars, and the ID of each star from
\citet{cud85} in order to make comparison with other studies easier.  Figure
\ref{m71} shows the confirmed members that were observed by SEGUE on
M71's CMD. We have chosen to show two CMDs for M71 because showing the 
stars with spectroscopic observations in Figure \ref{m71cmdfid} would
detract from the membership information and the new fiducial presented there.

\subsection{NCG 7789}

\begin{figure}[h!]
\centering
\includegraphics[scale=0.3,angle=270]{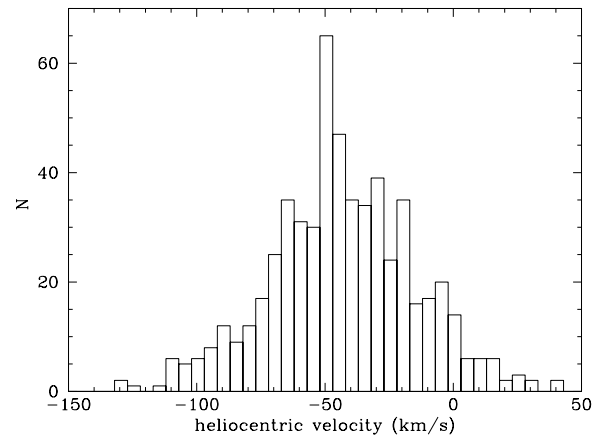}
\caption{Velocity histogram of stars on the NGC 7789 plate. The broad
  spread of velocities from foreground/background disk stars is
  clear. All our targeted members are in the highest peak at --50
  \kms, but roughly half the stars in this bin are still likely to be
  field stars. }
\label{7789vels}
\end{figure}

NGC 7789 is a populous open cluster with a metal abundance slightly
less than solar \citep{tau05} and age around 2 Gyr \citep{gim98}. We
selected targets using proper motion information from
\citet{mcnamara81} and velocities from \citet{gim2}.  Our selected
members have SEGUE radial velocities between --51.5 and --48.2 \kms.
Figure \ref{7789vels} shows the radial velocities of stars in the
SEGUE plate which observed NGC 7789, and illustrates some of the
problems of obtaining reliable member lists for open clusters. All of
the stars we targeted as likely giants are in the largest peak in the
histogram, centered on --50 \kms, but the contribution of foreground
and background disk stars to the velocity histogram, even at the exact
cluster velocity, is significant.  

\begin{figure}[h!]
\centering
\includegraphics[scale=0.45]{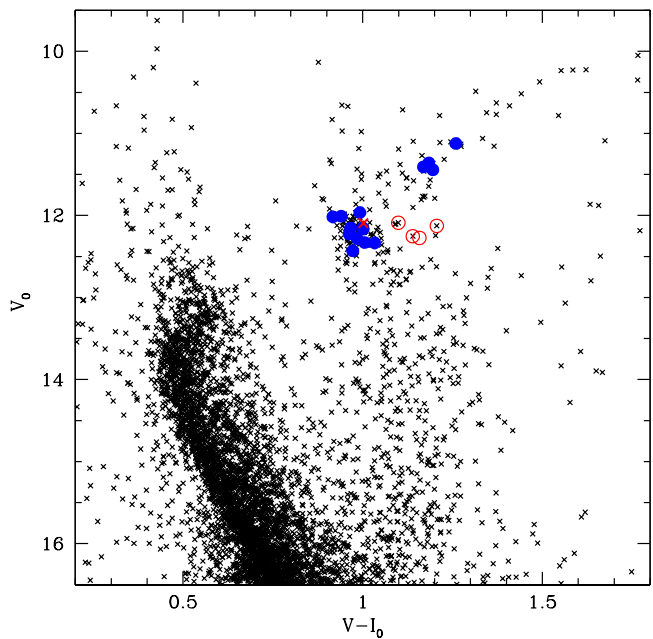}
\caption{VI CMD of the open cluster NGC 7789, based on data from
  \citet{gim98}. Stars which are proper motion, radial velocity (from
  SEGUE) and CMD members are shown as filled blue circles. Stars which
  are proper motion and radial velocity members which we chose to
  reject because of their position on the CMD are shown with red open
  circles, while the star which \protect \citet{gim2} note has radial velocity
  variations is shown with a red cross. }
\label{n7789}
\end{figure}

To produce our membership list, we rejected four of the stars
which were both proper motion and velocity members because of their
position in the color-magnitude diagram shown in Figure
\ref{n7789}. While these stars may be members whose variable reddening
moves them away from the cluster sequence, we have chosen to be
conservative and reject them, since a major aim of this paper is
simply to produce a collection of cluster stars which have a high likelihood 
of being members. We also rejected one star because \citep{gim2}  noted
that it had radial velocity variations.

 Since there are no $ugriz$ data available for NGC 7789, we show $V$
 and $I$ photometry from \citet{gim98} in the color-magnitude diagram
 of Figure \ref{n7789}, with the cluster members observed by SEGUE
 highlighted. The large contribution from foreground/background disk stars
 is clear in the CMD as well. The variable reddening can be seen in the scatter of
 colors and magnitudes in the clump in particular.  Individual
 estimates of reddening values were available for about one-third of
 the giants from the Vilnius photometry of \citet{vilnius7789}. We
 used these values where available, and their cluster value
 ($E(B-V)=0.25$) otherwise.
 
\begin{figure}[h!]
\centering
\includegraphics[scale=0.45]{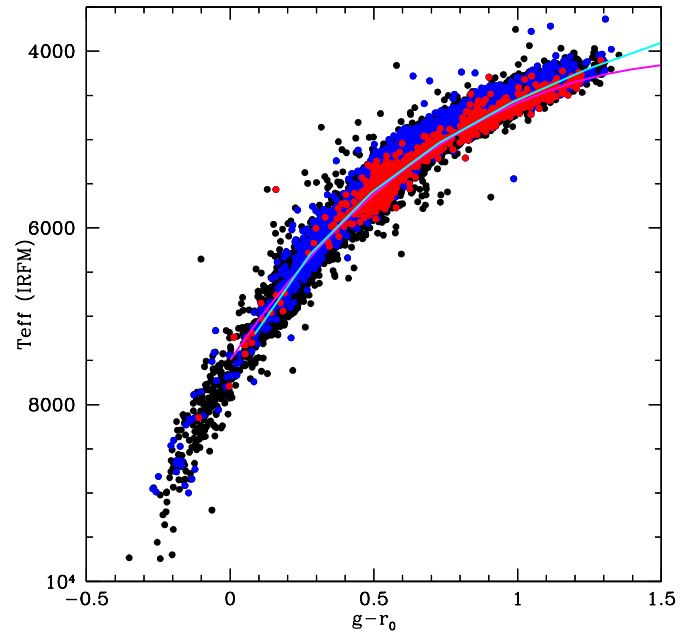}
\caption{\gr\ vs \tef\ calculated using the IRFM for stars with good
  \ug\ photometry and good 2MASS colors. Black points show all stars,
  while blue points show stars with $E(B-V)<0.025$, and red points
  show the data used to obtain the transformation between \tef\ and
  \gr\ for stars near solar metallicity, which have the additional
  restriction of $-0.2<[Fe/H]<0.25$ and \gr$>0$. Fit lines for
  equations (1) and (2) are shown with cyan and magenta lines
  respectively.  }
\label{teffcal}
\end{figure}

Since there are no \ug or \ugp data available for this cluster, we
transformed from $V-K_s$ to $g-r$ via \tef measurements.  First, we
used the relation between $V-K_s$ and \tef of \citet{ramirez}, the V
magnitudes of \citet{gim98}, 2MASS K magnitudes and $E(B-V)$ to derive
\tef\ for each star.  We chose to use $V-K_S$ because its relation
between effective temperature and color is the least sensitive to
[Fe/H] and gravity \citep{bessell}.   We then derived a relation
between \tef and $g-r$ using the Infrared Flux Method (IRFM hereafter) colors of stars observed with
SEGUE which had near-solar abundances. We selected 2068 stars with
SDSS spectra, good \ug, 2MASS J, H and K$_s$ colors, SSPP
metallicities between --0.2 and +0.25 (which have a mean metallicity
of --0.05, the same as NGC 7789) and low reddening: E(B-V) from
\citet{schlegel} less than 0.025.  Casagrande et al.\ (in preparation)
computed individual Infrared Flux Method \citep[see
][]{blackwell,luca10} temperatures for a large number of stars observed by
SEGUE, including these stars. Figure \ref{teffcal} shows the
relationship between \gr\ and this IRFM \tef\ for all stars with low
reddening (blue points) and for stars with near solar abundance (red
points). We fitted cubic relationships to these low-reddening, near
solar abundance stars as follows, first defining Q = 5040/\tef.

Q = 0.6728 + 0.4265 \gr  + 0.08841 \gr$^2$ -- 0.08881 \gr$^3$ \hfill (1)

The differences between this line and the actual \tef\ values have a sigma of 112 K.  

The inverse relation is:

\gr\ =  --0.3672 -- 0.7188 Q + 2.281 Q$^2$ -- 0.4688 Q$^3$ \hfill (2)

The differences between the fit line and the actual \gr\ values is
0.047 mag.  Assuming errors of 0.02 mag.\ for the V and $K_s$
magnitudes gives a random error estimate of 0.055 mag.\ for this entire
process.  Lastly, we added 100K to the \tef\ from \citet{ramirez}
to correct for the difference in absolute scale between their IRFM scale and
that used by Casagrande et al.\ \citep[see ][for a detailed discussion
  of this point]{luca10} then used equation (2) to calculate the value
of \gr\ for each star.

\begin{figure}[h!]
\centering
\includegraphics[scale=0.45]{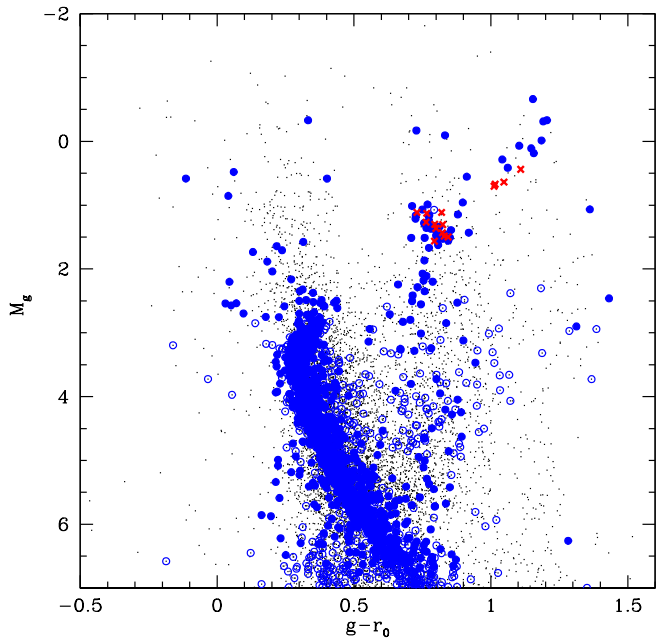}
\caption{CMD of NGC 6819, a near 'twin' of NGC 7789, with photometry
  from \citet{platais}.  Small black points show all stars in the
  field, open blue circles are stars with proper motion membership
  probability \protect \citep[from ][]{platais} greater than 50\%, and closed
  blue circles stars with membership probabilities greater than 80\%.
Red crosses show our NGC 7789 members. The close agreement in clump star 
colors indicates that our transformation from VI to $gr$ for NGC 7789 stars
is quite accurate.
}
\label{n6819}
\end{figure}

We checked this transformation using an open cluster which has a
similar age and metallicity to NGC 7789 but also has $g'r'$
observations \citep{platais}, NGC 6819. We used the standard
transformations of \citet{tucker} to transform the NGC 6819 photometry into
$g$ and $r$, and the cluster reddening of E(B-V)=0.16 \citep{ant14},
to make the CMD shown in Figure \ref{n6819}. Our 15 confirmed NGC 7789
red giants and clump stars are overplotted. The solid blue symbols are
proper motion members of NGC 6819, while the red crosses are our NGC
7789 members.  Since there are a range of cluster distance moduli for
NGC 6819 in the literature, we adjusted the NGC 6819 absolute
magnitudes until the clumps of the two clusters coincided. The
agreement of $g-r$ color between the two clumps is gratifying,
suggesting that our transformation from VI to $g-r$ is quite accurate.

Data on NGC 7789 members are listed in Table \ref{n7789mem}. The $V$
and $I$ photometry are from \citet{gim98}, radial velocities from our
SEGUE data and reddening values from \citet{vilnius7789}, and
\gr\ values from the transformation described above.

\subsection{NGC 6791}

\begin{figure}[h!]
\centering
\includegraphics[scale=0.45]{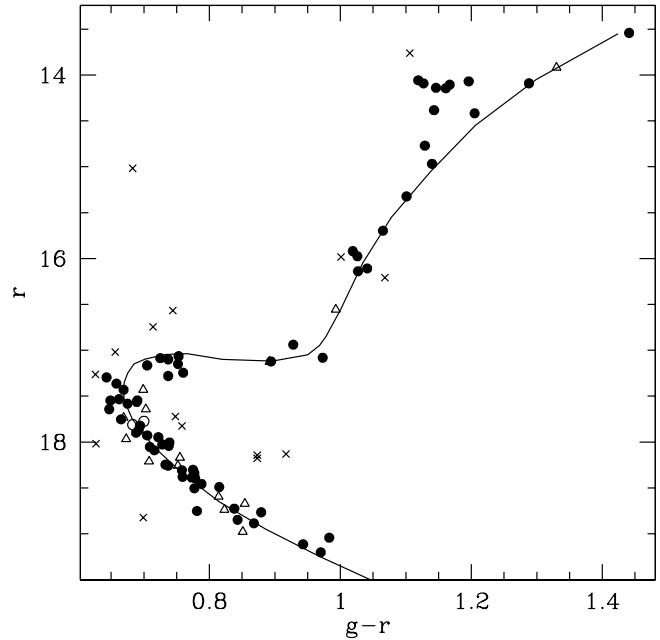}
\caption{$gr$ CMD of the rich open cluster NGC 6791, using data from
  \cite{an08}. Circles and triangles show radial velocity
  members. Stars with proper motion membership probabilities higher
  than 70\% are shown as filled circles, while those with proper
  motion membership probabilities smaller than 70\% are shown with
  open triangles. If no proper motion is available for the star, it is
  shown with an open circle. Crosses show stars classified as
  non-members because of their position in the CMD. }
\label{n6791}
\end{figure}

NGC 6791 is a particularly useful cluster because it anchors our
calibrations at the metal-rich end, having [Fe/H]= +0.39.  We used run 5416 of the photometry from \citet{an08} and applied
the ``ubercalibration'' corrections given in Table 1 of \citet{an13}.
To determine membership we used proper motion data from Cudworth
(private communication), choosing all stars with proper motion
membership probability greater than 70\%, velocities between --60 and
--48 km s$^{-1}$, and finally, by removing several stars whose
position on the CMD was not consistent with membership of the cluster.
We remind the reader that our aim here is to obtain a collection
of the most likely cluster members, not a complete set, and such rejection
is a conservative step for this purpose.
Figure \ref{n6791} shows the NGC 6791 stars observed by SEGUE on the
cluster CMD, and Table \ref{n6791mem} lists the likely members of the
cluster.

\section{A Cluster with no Proper-Motion data: Berkeley 29}

Useful proper motion data are not available for the open cluster Be
29. However, the cluster was observed using the SDSS imager on the 2.5m
 telescope at the end of the SDSS-III imaging season in Jan 2009
(Aihara et al. 2011), and we include its $gr$ photometry in this
work. Following the procedure in An et al. (2008) used for the other
SDSS clusters, we reduced SDSS imaging frames using the
DAOPHOT/ALLFRAME suite of programs (Stetson 1987; Stetson 1994), and
converted DAOPHOT magnitudes into the SDSS asinh values (Lupton et
al. 1999) using photometric zero points, extinction coefficients, and
airmass values. We tied our cluster photometry to the ubercal system,
as for the other cluster photometry discussed in Section 3, by deriving
zero-point offsets between DAOPHOT and DR8 photometry in low density
fields near Be 29. We checked the adjusted DAOPHOT magnitudes in these
cluster flanking fields with more recent values in DR12 (Alam et
al. 2015), and found $0.008$~mag differences in each of the $g$ and $r$
passbands. The $gr$ CMD of Be 29 is shown in Figure 7: the red clump
can be seen near $g- r = 0.8$ and $r = 17$.

Be 29's stars were in general fainter than those in the other clusters
we studied, because we aimed to obtain unsaturated photometry of its
giant branch. This led to larger than average velocity errors on the
Be 29 stars.  Because Be 29's radial velocity is close to that of the
disk stars in the field, contamination from foreground/background disk
stars is a problem, and we do not have a very accurate radial velocity
to use for member identification.  However, as we showed in our
discussion of NGC 7789, even full velocity and proper motion data for
each star is not sufficient to identify members for open clusters. Our
strategy, therefore, was to adopt a wide velocity window (10--40 \kms)
and visually inspect the spectra of the stars within this window to
determine if they were giants (and thus likely to belong to Be 29) or
dwarfs (and so not cluster members given their colors and magnitudes).
The criteria we used for this visual inspection included the strengths
of the Mg$b$/H feature near 5170\AA, the strengths of the CaI line at
4227\AA\ \citep[both described in detail in][]{spag6}, and the
relative strength of the SrII line at 4077\AA\ to the nearby FeI lines
at 4045 and 4063\AA\ \citep{rose84}.

\begin{figure}[h!]
\centering
\includegraphics[scale=0.45]{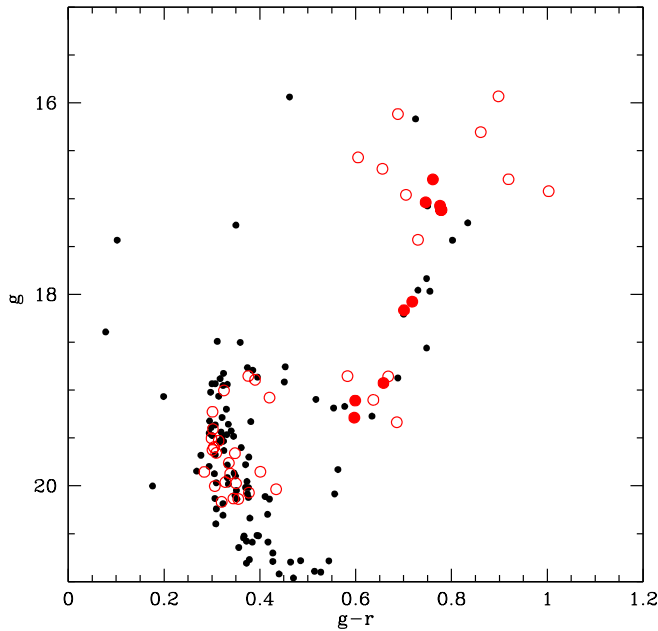}
\caption{CMD of the open cluster Be 29. Filled black circles show
  stars from the DAOphot reductions of the SDSS photometry which are
  within 3 arcmin radius of the cluster center. The most luminous
  giants in this plot are on the red horizontal branch. Open red
  circles show radial velocity members in the giant branch color
  range, while filled red circles show stars confirmed
  spectroscopically to be giants. It can be seen that there is significant
contamination by non-cluster members in the CMD. }
\label{be29}
\end{figure}

Table \ref{be29tab} provides detail on the Be 29 members observed by
SEGUE.

\section{Summary}

We present CMDs and membership information for stars from the
globular clusters M92, M13 and M71 and the open clusters Berkeley 29,
NGC 7789 and NGC 6791. 

For the nearby globular clusters M92 and M13 (\fe=--2.4 and --1.6
respectively), we describe the procedures used to observe the
brightest giants spectroscopically on the SDSS 2.5m and the
transformation of the \ugp\ magnitudes of these bright stars to
\ug. We used multiple criteria to isolate cluster members: the SEGUE
radial velocity, proper motions and CMD position.  We tabulate $griz$
magnitudes and SEGUE radial velocity measures for the 79 M92 members
and the 146 M13 members.

In the disk globular cluster M71 (\fe=--0.8), reddening is variable
due to its low galactic latitude. We mapped this variation using the
position of the main sequence turnoff in the photometry of
\citet{clem08} and give a table of variable reddening values for this
cluster. The improved reddening estimates produced a much cleaner CMD
and allowed us identify possible AGB members and to construct an
improved fiducial for the red giant branch, important for our K giant
distance estimates. This fiducial is tabulated in \citet{xue14}. We
give $gr$ photometry, reddening offsets and velocities for the 9 members
on the giant or horizontal branch which were observed by SEGUE.

Open clusters are traditionally harder to study because their low
galactic latitudes and low concentration lead to large contamination
from non-members, even when accurate radial velocity and proper
motions are available.

The open cluster Be 29 (\fe=--0.4) has no proper motion data
available, and because its stars are relatively faint, the SEGUE
radial velocities have errors of order 10 \kms. Thus we supplemented
the velocity and CMD position criteria with a visual inspection of the
SEGUE spectra in order to cull out the foreground dwarf stars in the
field of this cluster. We tabulate $gr$ photometry and velocity
measurements for the 10 cluster members on the lower giant branch and
red clump for this cluster.

The open cluster NGC 7789 (\fe=0.0) has both radial velocity and
proper motion data available, but no \ug\ or \ugp\ data. We derived a
transformation from V-K to $gr$ for stars with near-solar abundances
via recent \tef\ calibrations of stars with good SDSS photometry and
SEGUE spectra. We validated this transformation via CMD comparisons
with the very similar open cluster NGC 6819, and give photometry and
velocities for the 15 giant branch or red clump members of NGC 7789
observed by SEGUE.

The super-metal-rich cluster NGC 6791 (\fe=+0.4) has higher concentration than many
open clusters, so the combination of CMD position, radial velocity and proper
motion is sufficient to identify cluster members. We tabulate $griz$ photometry
and SEGUE velocities for the 67 cluster members observed by SEGUE.

The information presented in this paper will be useful in calibrations
and tests of the SEGUE observations, particularly for red giant stars,
and will be used in the series of papers titled ``The SEGUE K giant
survey''.  These papers include a detailed description of the
technique of identifying K giants in SEGUE data (Morrison et al, in
preparation), the calculation of distances for the $\sim$6,000 K
giants identified by the survey \citep{xue14}, a study of substructure
in halo giants out to 100 kpc \citep{bill}, the study of the radial
profile of the halo out to 60 kpc, which also includes an estimate of
the (slight) radial metallicity gradient in the halo \citep{xue15} and
two papers currently in preparation, one on the halo MDF and its
variation with distance, and the other on the [$\alpha$/Fe] ratios of
$\sim$2000 K giants in the halo.

\section{Acknowledgements}

We thank Tad Pryor for kindly sharing his unpublished velocity data
for M71 with us, and Bruce Twarog for suggesting that we use NGC 6819
to help transform the NGC 7789 data to $gr$. We also thank the anonymous
referee for a very helpful report. This research used the
facilities of the Canadian Astronomy Data Centre operated by the
National Research Council of Canada with the support of the Canadian
Space Agency. This publication also makes use of data products from
the Two Micron All Sky Survey, which is a joint project of the
University of Massachusetts and the Infrared Processing and Analysis
Center/California Institute of Technology, funded by the National
Aeronautics and Space Administration and the National Science
Foundation.  This work was supported in part by the National Science
Foundation under Grant No. PHYS-1066293 and the hospitality of the
Aspen Center for Physics, and by grants AST-1009886 to HLM and AST-121989 to
HLM, PH and CR. TCB acknowledges partial support for this work by grant PHY
08-22648: Physics Frontiers Center/Joint Institute for Nuclear
Astrophysics (JINA), awarded by the U.S. National Science Foundation.

Funding for SDSS-III has been provided by the Alfred P. Sloan
Foundation, the Participating Institutions, the National Science
Foundation, and the U.S. Department of Energy Office of Science. The
SDSS-III web site is http://www.sdss3.org/.  SDSS-III is managed by
the Astrophysical Research Consortium for the Participating
Institutions of the SDSS-III Collaboration including the University of
Arizona, the Brazilian Participation Group, Brookhaven National
Laboratory, Carnegie Mellon University, University of Florida, the
French Participation Group, the German Participation Group, Harvard
University, the Instituto de Astrofisica de Canarias, the Michigan
State/Notre Dame/JINA Participation Group, Johns Hopkins University,
Lawrence Berkeley National Laboratory, Max Planck Institute for
Astrophysics, Max Planck Institute for Extraterrestrial Physics, New
Mexico State University, New York University, Ohio State University,
Pennsylvania State University, University of Portsmouth, Princeton
University, the Spanish Participation Group, University of Tokyo,
University of Utah, Vanderbilt University, University of Virginia,
University of Washington, and Yale University.

\clearpage

\LongTables

\begin{deluxetable*}{llrrccr}
\tablewidth{0pt}
\tabletypesize{\tiny} 
\tablecaption{Cluster Properties \label{clusterdata}} 
\tablehead{
  \colhead{ID} &
  \colhead{Alternate name} &
  \colhead{$l(^{\circ})$\tablenotemark{a}} &
  \colhead{$b(^{\circ})$\tablenotemark{a}} &
  \colhead{E(B--V)}  &
  \colhead{$(m-M)_0$}  &
  \colhead{\fe}
}
\startdata
 M92   & NGC 6341 &  $ 68.34$ & $+34.86$ & $0.02$\tablenotemark{e} &  $14.64$\tablenotemark{d} & $-2.38$\tablenotemark{e}    \\
 M13   & NGC 6205 &  $ 59.01$ & $+40.91$ & $0.02$\tablenotemark{e} &  $14.38$\tablenotemark{d} & $-1.60$\tablenotemark{e}    \\
 M71   & NGC 6838 &  $ 56.74$ & $ -4.56$ & $0.28$\tablenotemark{c} &  $12.86$\tablenotemark{c} & $-0.81$\tablenotemark{e}    \\
 Be 29\tablenotemark{k} &  & $197.95$ & $ +7.98$ & $0.08$ &  $15.6 $                  & $-0.38$\\
 NGC 7789 & & $115.48$ & $ -5.37$ & $0.25$\tablenotemark{i} &  $11.33$\tablenotemark{j} & $-0.04$\tablenotemark{k}    \\
  NGC 6791  & & $ 69.96$ & $+10.90$ & $0.16$\tablenotemark{b} &  $13.01$\tablenotemark{b} & $ 0.39$\tablenotemark{f}    \\
 \enddata
\tablenotetext{a}{The coordinates are based on the compilation of \citet{an08} except for NGC 7789 and Berkeley 29.}
\tablenotetext{b}{\citet{bro11}; $(m-M)_0$ is based on $(m-M)_V$ assuming $A_V=3.1*$\ebv.}
\tablenotetext{c}{\citet{gru02}; $(m-M)_0$ derived from  Hipparcos \citep{per97} subdwarf fitting.}
\tablenotetext{d}{\citet{car00}; $(m-M)_0$ derived from  Hipparcos subdwarf fitting.}
\tablenotetext{e}{\citet{kraftivans03}; their globular cluster metallicity scale is based on the FeII lines from high-resolution spectra of giants.}
\tablenotetext{f}{We averaged the \fe\ measurements of  \citet{pet98},  \citet{gra06},  \citet{car06} and
\citet{bro11} (+0.40, +0.47,  +0.39 and  +0.29 respectively).}
\tablenotetext{g}{\citet{ant06}}
\tablenotetext{h}{\citet{jacobson11}}
\tablenotetext{i}{\citet{vilnius7789}}
\tablenotetext{j}{\citet{tau05}}
\tablenotetext{k}{Reddening is from \citet{car04}, $(m-M)_0$ is from \citet{ses08} and \fe~ is the average of the \citet{car04} and \citet{ses08} values.}
\tablenotetext{l}{\citet{taylor}}
\tablenotetext{m}{\citet{an07}}
\tablenotetext{n}{\citet{onehag}}

\end{deluxetable*}

\clearpage

\begin{deluxetable*}{llrcl}
\tablewidth{0pt}
\tablecolumns{5}
\tabletypesize{\tiny}
\tablecaption{Plate Information for Clusters in this Paper \label{clusterplates}}
\tablehead{\colhead{Cluster ID} & \colhead{Alternate} & \colhead{Plate} &
  \colhead{MJD} & \colhead{Plate type}  \\
 }
\startdata
 M92 & NGC 6341 & 2247 & 53857 & Offset \\     
 M92 & NGC 6341 & 2247 & 54169 & Bright \\     
 M92 & NGC 6341 & 2256 & 53859 & Faint \\   \\   
                                               
 M13 & NGC 6205 & 2255 & 53565 & Offset \\     
 M13 & NGC 6205 & 2255 & 53565 & Very Bright \\
 M13 & NGC 6205 & 2174 & 53521 & Bright \\     
 M13 & NGC 6205 & 2185 & 53532 & Faint \\ \\      
                 
 M71 & NGC 6838 & 2333 & 53682 & \\
 M71 & NGC 6838 & 2338 & 53679 & \\ \\
                 
 Be 29   &      & 3334 & 54927 &  \\
 Be 29   &      & 3335 & 54922 & \\ \\

NGC 7789 && 2337 & 53991 & Bright \\ \\
                 
NGC 6791 && 2800 & 54326 & Bright \\
NGC 6791 && 2821 & 54393 & Faint \\
 
\enddata 
\end{deluxetable*}

\begin{deluxetable*}{rrrrrrrrrrrrc }
\tablewidth{0pt}
\tablecolumns{13}
\tabletypesize{\tiny}
\tablecaption{M92 Members Observed by SEGUE \label{m92mem}}
\tablehead{\colhead{Plate} & \colhead{MJD} & \colhead{Fiber} &
  \colhead{$r$} & \colhead{error} &\colhead{$g-r$} &\colhead{$g-i$} &\colhead{$g-z$} &
\colhead{RA } &\colhead{Dec} &\colhead{Velocity} &\colhead{Phot.} &\colhead{PM prob} \\
 &  &  &
 (mag.)  & (mag.) & (mag.) &(mag.) &(mag.) &\colhead{(2000)} &\colhead{ (2000)} &\colhead{(km s$^{-1}$)} &\colhead{ref.} &\colhead{(\%)} \\
} 
\startdata

2247 & 53857 &  402 &    14.456  &   0.006  &    0.621  &    0.893  &    1.040  &   259.1997  &    43.1044  &   --114.6 &  1 &  99   \\
2247 & 53857 &  403 &    14.218  &   0.006  &    0.607  &    0.884  &    1.047  &   259.1831  &    43.1255  &   --103.4 &  1 &  99   \\
2247 & 53857 &  412 &    13.410  &   0.002  &    0.656  &    0.963  &    1.148  &   259.1429  &    43.1268  &   --113.3 &  1 &  99   \\
2247 & 53857 &  416 &    13.448  &   0.003  &    0.642  &    0.948  &    1.128  &   259.1572  &    43.1448  &   --122.2 &  1 &  99   \\
2247 & 53857 &  444 &    13.592  &   0.003  &    0.636  &    0.943  &    1.116  &   259.2514  &    43.1966  &   --109.5 &  1 &  99   \\ 
2247 & 53857 &  449 &    13.226  &   0.002  &    0.723  &    1.065  &    1.234  &   259.2452  &    43.2532  &   --107.9 &  1 &  99   \\
2247 & 53857 &  453 &    14.333  &   0.038  &    0.583  &    0.842  &    0.976  &   259.2280  &    43.1740  &   --110.0 &  1 &  99   \\
2247 & 53857 &  455 &    12.462  &   0.004  &    0.857  &    1.252  &    1.454  &   259.2078  &    43.1781  &   --115.4 &  1 &  99   \\
2247 & 53857 &  458 &    13.109  &   0.002  &    0.752  &    1.107  &    1.276  &   259.2406  &    43.2365  &   --118.0 &  1 &  99   \\
2247 & 53857 &  460 &    13.813  &   0.011  &    0.634  &    0.969  &    1.142  &   259.2313  &    43.0843  &   --120.1 &  1 &  99   \\
2247 & 53857 &  486 &    14.586  &   0.005  &    0.596  &    0.865  &    1.027  &   259.1491  &    42.9443  &   --118.5 &  1 &  \nodata  \\ 
2247 & 53857 &  514 &    14.344  &   0.003  &    0.596  &    0.877  &    1.028  &   259.2487  &    43.0183  &   --119.8 &  1 &  99   \\
2247 & 53857 &  516 &    14.285  &   0.005  &    0.420  &    0.614  &    0.684  &   259.2202  &    43.0582  &   --110.1 &  1 &  99   \\
2247 & 53857 &  522 &    11.619  &   0.006  &    1.159  &    1.659  &    1.930  &   259.3405  &    43.2149  &   --103.4 &  1 &  99   \\
2247 & 53857 &  523 &    12.740  &   0.003  &    0.809  &    1.176  &    1.360  &   259.3189  &    43.1792  &   --104.6 &  1 &  99   \\
2247 & 53857 &  525 &    14.024  &   0.014  &    0.658  &    0.933  &    1.095  &   259.3403  &    43.1842  &   --110.7 &  1 &  99   \\
2247 & 53857 &  526 &    11.961  &   0.011  &    0.970  &    1.441  &    1.702  &   259.2935  &    43.1855  &   --106.2 &  1 &  99   \\
2247 & 53857 &  529 &    13.895  &   0.005  &    0.556  &    0.824  &    0.980  &   259.3320  &    43.2451  &   --112.3 &  1 &  99   \\
2247 & 53857 &  532 &    14.305  &   0.009  &    0.476  &    0.691  &    0.818  &   259.3376  &    43.1035  &   --108.7 &  1 &  99   \\
2247 & 53857 &  537 &    14.332  &   0.006  &    0.609  &    0.887  &    1.032  &   259.3146  &    43.0831  &   --113.3 &  1 &  99   \\
2247 & 53857 &  553 &    14.833  &   0.001  &    0.341  &    0.486  &    0.583  &   259.3048  &    43.0001  &   --102.9 &  1 &  99       \\
2247 & 53857 &  559 &    14.371  &   0.005  &    0.595  &    0.875  &    1.033  &   259.2664  &    43.0341  &   --110.2 &  1 &  99   \\
2247 & 53857 &  572 &    12.407  &   0.003  &    0.856  &    1.230  &    1.459  &   259.3821  &    43.0949  &   --107.3 &  1 &  99   \\
2247 & 53857 &  576 &    14.507  &   0.010  &    0.599  &    0.879  &    1.035  &   259.3699  &    43.1675  &   --114.0 &  1 &  99   \\
2247 & 53857 &  577 &    14.691  &   0.004  &    0.571  &    0.831  &    0.982  &   259.3709  &    43.1153  &   --114.1 &  1 &  99       \\
2247 & 53857 &  580 &    14.353  &   0.009  &    0.598  &    0.883  &    1.020  &   259.3730  &    43.2041  &   --113.7 &  1 &  99   \\
2247 & 53857 &  583 &    13.489  &   0.005  &    0.690  &    1.004  &    1.200  &   259.3425  &    43.0805  &   --112.8 &  1 &  99   \\
2247 & 53857 &  609 &    14.660  &   0.005  &    0.606  &    0.871  &    0.998 &    259.5197  &    43.1712  &   --118.5 &  2 &  84     \\
2247 & 53857 &  612 &    14.889  &   0.002  &    0.566  &    0.832  &    0.977  &   259.4598  &    43.2295  &   --105.3 &  1 &  \nodata  \\ 
2247 & 54169 &  361 &    15.583  &   0.004  &   -0.225  &   -0.395  &   -0.463  &   259.0527  &    43.1739  &   --115.1 &  1 &  \nodata  \\ 
2247 & 54169 &  380 &    15.664  &   0.002  &    0.522  &    0.749  &    0.904  &   259.1245  &    43.1009  &   --109.5 &  1 &  \nodata  \\ 
2247 & 54169 &  408 &    14.865  &   0.007  &    0.567  &    0.816  &    0.973  &   259.1516  &    43.1156  &   --120.0 &  1 &  99       \\
2247 & 54169 &  418 &    15.063  &   0.003  &    0.570  &    0.810  &    0.959  &   259.1925  &    43.0829  &   --119.6 &  1 &  99       \\
2247 & 54169 &  441 &    15.760  &   0.002  &   -0.316  &   -0.533  &   -0.641  &   259.2120  &    43.1897  &   --115.4 &  1 &  \nodata  \\ 
2247 & 54169 &  444 &    15.590  &   0.003  &    0.499  &    0.740  &    0.871  &   259.1783  &    43.2465  &   --112.3 &  1 &  \nodata  \\ 
2247 & 54169 &  449 &    15.015  &   0.003  &    0.549  &    0.800  &    0.946  &   259.2012  &    43.1713  &   --116.5 &  1 &  \nodata  \\ 
2247 & 54169 &  451 &    15.753  &   0.003  &    0.527  &    0.744  &    0.888  &   259.2681  &    43.0696  &   --121.9 &  1 &  \nodata  \\ 
2247 & 54169 &  452 &    15.565  &   0.005  &    0.495  &    0.746  &    0.897  &   259.1898  &    43.2296  &   --121.4 &  1 &  \nodata  \\ 
2247 & 54169 &  484 &    14.586  &   0.005  &    0.596  &    0.865  &    1.027  &   259.1490  &    42.9443  &   --115.0 &  1 &  \nodata  \\  
2247 & 54169 &  504 &    14.638  &   0.022  &    0.554  &    0.839  &    1.002  &   259.3471  &    42.9488  &   --112.1 &  1 &  \nodata  \\ 
2247 & 54169 &  521 &    15.651  &   0.004  &   -0.273  &   -0.454  &   -0.545  &   259.2528  &    43.2175  &   --119.3 &  1 &  99       \\
2247 & 54169 &  526 &    15.592  &   0.004  &   -0.257  &   -0.447  &   -0.527  &   259.2901  &    43.0796  &   --117.6 &  1 &  99       \\
2247 & 54169 &  531 &    16.043  &   0.005  &    0.511  &    0.724  &    0.830 &    259.3130  &    43.2645  &   --115.2 &  2 &  \nodata \\
2247 & 54169 &  538 &    17.658  &   0.007  &    0.373  &    0.525  &    0.583 &    259.3413  &    43.2580  &   --107.6 &  2 &  \nodata \\
2247 & 54169 &  549 &    15.237  &   0.006  &   -0.199  &   -0.343  &   -0.381  &   259.4017  &    43.0199  &   --121.2 &  1 &  97       \\
2247 & 54169 &  563 &    14.989  &   0.006  &    0.568  &    0.816  &    0.955  &   259.3296  &    43.2152  &   --111.7 &  1 &  99       \\
2247 & 54169 &  565 &    15.910  &   0.008  &   -0.335  &   -0.579  &   -0.701  &   259.3446  &    43.1587  &   --109.7 &  1 &  \nodata   \\
2247 & 54169 &  567 &    17.255  &   0.005  &    0.453  &    0.595  &    0.626 &    259.4499  &    43.3070  &   --118.4 &  2 &  \nodata \\
2247 & 54169 &  568 &    15.370  &   0.006  &   -0.181  &   -0.314  &   -0.357  &   259.3660  &    43.1475  &   --113.8 &  1 &  99       \\
2247 & 54169 &  573 &    15.604  &   0.002  &    0.492  &    0.689  &    0.796  &   259.3214  &    43.0742  &   --116.3 &  1 &  \nodata  \\ 
2247 & 54169 &  575 &    17.519  &   0.006  &    0.427  &    0.587  &    0.654 &    259.3812  &    43.2469  &   --111.4 &  2 &  \nodata \\
2247 & 54169 &  581 &    15.963  &   0.007  &    0.495  &    0.711  &    0.837  &   259.3938  &    43.0711  &   --110.5 &  1 &  \nodata   \\
2247 & 54169 &  584 &    17.419  &   0.006  &    0.434  &    0.615  &    0.704 &    259.4844  &    43.0595  &   --103.4 &  2 &  \nodata \\
2247 & 54169 &  589 &    16.587  &   0.006  &    0.481  &    0.675  &    0.802 &    259.4322  &    43.0634  &   --114.7 &  2 &  \nodata \\
2247 & 54169 &  601 &    16.062  &   0.005  &    0.517  &    0.729  &    0.817 &    259.5772  &    43.1990  &   --112.1 &  2 &  \nodata \\
2247 & 54169 &  608 &    14.889  &   0.002  &    0.566  &    0.832  &    0.977  &   259.4598  &    43.2295  &   --120.3 &  1 &  \nodata  \\ 
2247 & 54169 &  610 &    14.660  &   0.005  &    0.606  &    0.871  &    0.998 &    259.5197  &    43.1712  &   --117.2 &  2 &  \nodata \\
2247 & 54169 &  612 &    15.999  &   0.007  &   -0.308  &   -0.527  &   -0.636  &   259.4042  &    43.1310  &   --116.4 &  1 &  \nodata   \\
2247 & 54169 &  616 &    17.522  &   0.005  &    0.433  &    0.598  &    0.668 &    259.3905  &    43.1896  &   --115.4 &  2 &  \nodata \\
2247 & 54169 &  620 &    17.097  &   0.005  &    0.473  &    0.663  &    0.741 &    259.4374  &    43.1356  &   --103.9 &  2 &  \nodata \\
2256 & 53859 &  521 &    18.947  &   0.012  &    0.259  &    0.361  &    0.297 &    259.3361  &    43.2903  &   --117.3 &  2 &  \nodata \\
2256 & 53859 &  525 &    18.285  &   0.009  &    0.239  &    0.343  &    0.406 &    259.3690  &    43.2710  &   --111.2 &  2 &  \nodata \\
2256 & 53859 &  534 &    19.172  &   0.014  &    0.279  &    0.337  &    0.335 &    259.3679  &    43.2307  &   --104.4 &  2 &  \nodata \\
2256 & 53859 &  537 &    19.080  &   0.014  &    0.279  &    0.369  &    0.416 &    259.4246  &    43.1206  &   --112.4 &  2 &  \nodata \\
2256 & 53859 &  537 &    20.197  &   0.036  &    0.525  &    0.708  &    0.760 &    259.4246  &    43.1206  &   --112.4 &  2 &  \nodata \\
2256 & 53859 &  538 &    17.373  &   0.006  &    0.453  &    0.625  &    0.688 &    259.3800  &    43.2105  &   --120.7 &  2 &  \nodata \\
2256 & 53859 &  539 &    18.661  &   0.011  &    0.277  &    0.405  &    0.452 &    259.3406  &    43.2491  &   --113.1 &  2 &  \nodata \\
2256 & 53859 &  561 &    18.975  &   0.013  &    0.262  &    0.359  &    0.367 &    259.4673  &    43.1504  &   --112.9 &  2 &  \nodata \\
2256 & 53859 &  562 &    18.709  &   0.012  &    0.232  &    0.323  &    0.270 &    259.3947  &    43.2326  &   --119.8 &  2 &  \nodata \\
2256 & 53859 &  563 &    19.064  &   0.014  &    0.287  &    0.356  &    0.315 &    259.4192  &    43.3309  &   --111.3 &  2 &  \nodata \\
2256 & 53859 &  567 &    19.434  &   0.016  &    0.332  &    0.407  &    0.459 &    259.4300  &    43.3105  &   --103.5 &  2 &  \nodata \\
2256 & 53859 &  569 &    19.511  &   0.017  &    0.294  &    0.427  &    0.480 &    259.4556  &    43.1986  &   --109.2 &  2 &  \nodata \\
2256 & 53859 &  576 &    18.356  &   0.008  &    0.258  &    0.320  &    0.334 &    259.4356  &    43.1723  &   --109.1 &  2 &  \nodata \\
2256 & 53859 &  578 &    19.320  &   0.021  &    0.362  &    0.458  &    0.416 &    259.4461  &    43.1315  &   --103.3 &  2 &  \nodata \\
2256 & 53859 &  604 &    18.693  &   0.010  &    0.260  &    0.349  &    0.385 &    259.5561  &    43.2827  &   --121.2 &  2 &  \nodata \\
2256 & 53859 &  608 &    17.984  &   0.008  &    0.302  &    0.393  &    0.419 &    259.4972  &    43.3424  &   --122.2 &  2 &  \nodata \\
2256 & 53859 &  612 &    18.313  &   0.010  &    0.253  &    0.345  &    0.332 &    259.4838  &    43.2025  &   --115.4 &  2 &  \nodata \\
2256 & 53859 &  616 &    18.628  &   0.009  &    0.279  &    0.320  &    0.273 &    259.6061  &    43.1568  &   --117.3 &  2 &  \nodata \\
2256 & 53859 &  621 &    17.672  &   0.005  &    0.408  &    0.560  &    0.610 &    259.6847  &    43.0911  &   --112.6 &  2 &  \nodata \\

\enddata

\tablerefs{1. \citet{clem08}, 2. \citet{an08}}

\end{deluxetable*}

\begin{deluxetable*}{rrrrrrrrrrrrc}
\tablewidth{0pt}
\tablecolumns{13}
\tabletypesize{\tiny}
\tablecaption{M13 Members Observed by SEGUE \label{m13tab}}
\tablehead{\colhead{Plate} & \colhead{MJD} & \colhead{Fiber} &
  \colhead{$r$} & \colhead{error} &\colhead{$g-r$} &\colhead{$g-i$} &\colhead{$g-z$} &
\colhead{RA } &\colhead{Dec} &\colhead{Velocity} &\colhead{Phot.} &\colhead{PM prob} \\
 &  &  &
  \colhead{ (mag.)} & \colhead{(mag.)} &\colhead{(mag.)} &\colhead{(mag.)} &\colhead{(mag.)} &\colhead{(2000)} &\colhead{(2000)} &\colhead{(km s$^{-1}$)} &\colhead{ref.} &\colhead{(\%)} \\
} 
\startdata

 2174&  53521 &   128&  17.060& 0.013&   0.471&   0.682&   0.753 & 250.4742 &  36.3098& -249.0 &  2& \nodata \\
 2174&  53521 &   133&  16.421& 0.010&   0.508&   0.728&   0.813 & 250.5125 &  36.3211& -242.6 &  2& \nodata \\
 2174&  53521 &   153&  15.124& 0.020&  -0.159&  -0.304&  -0.347 & 250.4664 &  36.4093& -247.8 &  2&  99	   \\
 2174&  53521 &   406&  17.522& 0.011&   0.430&   0.610&   0.678 & 250.2975 &  36.6566& -249.8 &  2& \nodata \\
 2174&  53521 &   444&  14.532& 0.005&   0.621&   0.913&   1.029 & 250.3637 &  36.5395& -243.7 &  2&  99	   \\
 2174&  53521 &   445&  17.523& 0.010&   0.425&   0.600&   0.661 & 250.3154 &  36.5818& -247.6 &  2& \nodata \\
 2174&  53521 &   447&  17.276& 0.009&   0.466&   0.658&   0.708 & 250.3488 &  36.6371& -246.1 &  2& \nodata \\
 2174&  53521 &   456&  16.942& 0.009&   0.486&   0.677&   0.761 & 250.3559 &  36.6084& -248.2 &  2& \nodata \\
 2174&  53521 &   461&  16.926& 0.008&   0.488&   0.686&   0.780 & 250.4162 &  36.5927& -245.1 &  2& \nodata \\
 2174&  53521 &   480&  16.441& 0.005&   0.511&   0.735&   0.823 & 250.3776 &  36.5606& -244.9 &  2& \nodata \\
 2174&  53521 &   554&  15.314& 0.010&   0.585&   0.826&   0.949 & 250.4525 &  36.7311& -248.1 &  2& \nodata \\
2174 &  53521 &   121&  15.767 &  0.004 &  0.522 &  0.754 &  0.867 & 250.5334 &  36.3239 & -249.4 &   1 &  \nodata \\
2174 &  53521 &   131&  14.601 &  0.022 &  0.595 &  0.878 &  1.044 & 250.4894 &  36.3321 & -243.0 &   1 &  99 \\
2174 &  53521 &   136&  15.511 &  0.004 & -0.286 & -0.468 & -0.588 & 250.4906 &  36.3635 & -242.9 &   1 &  98 \\
2174 &  53521 &   145&  14.357 &  0.002 &  0.658 &  0.930 &  1.063 & 250.4505 &  36.3933 & -246.4 &   1 &  99 \\
2174 &  53521 &   154&  14.475 &  0.003 &  0.636 &  0.887 &  1.056 & 250.4662 &  36.3263 & -249.9 &   1 &  99 \\
2174 &  53521 &   156&  15.142 &  0.006 &  0.577 &  0.830 &  0.946 & 250.3522 &  36.4095 & -239.7 &   1 &  99 \\
2174 &  53521 &   157&  15.412 &  0.004 & -0.254 & -0.440 & -0.548 & 250.4520 &  36.3018 & -247.2 &   1 &  99 \\
2174 &  53521 &   158&  15.112 &  0.004 &  0.586 &  0.822 &  0.957 & 250.4085 &  36.3039 & -243.9 &   1 &  99 \\
2174 &  53521 &   167&  14.255 &  0.003 &  0.654 &  0.932 &  1.089 & 250.2756 &  36.4229 & -246.5 &   1 &  99 \\
2174 &  53521 &   168&  14.721 &  0.003 &  0.616 &  0.874 &  1.023 & 250.2608 &  36.4377 & -244.0 &   1 &  99 \\
2174 &  53521 &   171&  14.367 &  0.003 &  0.656 &  0.935 &  1.083 & 250.3129 &  36.3983 & -247.0 &   1 &  90 \\
2174 &  53521 &   172&  14.753 &  0.003 &  0.619 &  0.885 &  1.026 & 250.3078 &  36.4174 & -247.8 &   1 &  99 \\
2174 &  53521 &   176&  15.232 &  0.002 &  0.577 &  0.824 &  0.919 & 250.3261 &  36.3471 & -240.3 &   1 &  99 \\
2174 &  53521 &   412&  17.085 &  0.004 &  0.464 &  0.657 &  0.749 & 250.2389 &  36.5871 & -242.8 &   1 &  \nodata \\
2174 &  53521 &   414&  17.060 &  0.003 &  0.465 &  0.672 &  0.776 & 250.2673 &  36.5864 & -241.9 &   1 &  \nodata \\
2174 &  53521 &   418&  15.919 &  0.003 &  0.547 &  0.775 &  0.923 & 250.2429 &  36.7086 & -247.4 &   1 &  \nodata \\
2174 &  53521 &   442&  15.116 &  0.002 & -0.187 & -0.328 & -0.421 & 250.3339 &  36.6145 & -249.2 &   1 &  99 \\
2174 &  53521 &   449&  14.806 &  0.002 &  0.604 &  0.865 &  1.036 & 250.3311 &  36.5077 & -244.1 &   1 &  99 \\
2174 &  53521 &   457&  14.620 &  0.003 &  0.612 &  0.878 &  0.974 & 250.3618 &  36.4246 & -244.6 &   1 &  99 \\
2174 &  53521 &   458&  14.511 &  0.002 &  0.632 &  0.902 &  1.060 & 250.3154 &  36.4639 & -245.9 &   1 &  99 \\
2174 &  53521 &   459&  14.665 &  0.001 &  0.611 &  0.886 &  1.050 & 250.3160 &  36.5549 & -245.0 &   1 &  99 \\
2174 &  53521 &   460&  14.246 &  0.002 &  0.648 &  0.932 &  1.105 & 250.3239 &  36.4916 & -238.9 &   1 &  99 \\
2174 &  53521 &   462&  14.560 &  0.002 &  0.623 &  0.911 &  1.056 & 250.3755 &  36.5912 & -245.5 &   1 &  99 \\
2174 &  53521 &   463&  15.115 &  0.001 &  0.595 &  0.867 &  0.999 & 250.4507 &  36.5948 & -243.8 &   1 &  99 \\
2174 &  53521 &   464&  15.163 &  0.003 & -0.207 & -0.331 & -0.411 & 250.3978 &  36.6046 & -239.1 &   1 &  99 \\
2174 &  53521 &   466&  14.919 &  0.002 & -0.018 & -0.059 & -0.091 & 250.4609 &  36.5551 & -241.7 &   1 &  99 \\
2174 &  53521 &   467&  14.620 &  0.003 &  0.614 &  0.888 &  1.071 & 250.3318 &  36.6897 & -242.9 &   1 &  \nodata \\
2174 &  53521 &   470&  14.258 &  0.007 &  0.649 &  0.941 &  1.100 & 250.3886 &  36.5412 & -241.2 &   1 &  99 \\
2174 &  53521 &   471&  15.802 &  0.002 &  0.541 &  0.777 &  0.909 & 250.4051 &  36.6807 & -248.8 &   1 &  \nodata \\
2174 &  53521 &   472&  14.888 &  0.004 & -0.004 & -0.047 & -0.119 & 250.4393 &  36.4306 & -243.1 &   1 &  99 \\
2174 &  53521 &   475&  14.333 &  0.002 &  0.649 &  0.926 &  1.086 & 250.4188 &  36.5268 & -239.4 &   1 &  99 \\
2174 &  53521 &   476&  15.207 &  0.002 &  0.593 &  0.840 &  0.968 & 250.4538 &  36.5347 & -249.7 &   1 &  99 \\
2174 &  53521 &   477&  14.629 &  0.002 &  0.642 &  0.904 &  1.057 & 250.4330 &  36.4116 & -240.9 &   1 &  99 \\
2174 &  53521 &   486&  15.466 &  0.003 & -0.273 & -0.477 & -0.572 & 250.4692 &  36.5147 & -245.9 &   1 &  99 \\
2174 &  53521 &   487&  14.507 &  0.001 &  0.651 &  0.910 &  1.078 & 250.5424 &  36.6308 & -249.8 &   1 &  98 \\
2174 &  53521 &   488&  15.197 &  0.004 &  0.589 &  0.842 &  0.956 & 250.5207 &  36.5268 & -246.1 &   1 &  99 \\
2174 &  53521 &   489&  15.037 &  0.003 &  0.599 &  0.855 &  1.000 & 250.5569 &  36.5541 & -246.5 &   1 &  99 \\
2174 &  53521 &   491&  15.463 &  0.002 & -0.253 & -0.440 & -0.553 & 250.5045 &  36.5626 & -249.7 &   1 &  99 \\
2174 &  53521 &   493&  14.519 &  0.008 &  0.633 &  0.906 &  1.022 & 250.4687 &  36.4504 & -247.9 &   1 &  99 \\
2174 &  53521 &   497&  14.836 &  0.002 &  0.629 &  0.887 &  1.037 & 250.5392 &  36.5664 & -240.3 &   1 &  99 \\
2174 &  53521 &   498&  14.558 &  0.002 &  0.627 &  0.897 &  1.043 & 250.5455 &  36.4092 & -240.7 &   1 &  99 \\
2174 &  53521 &   499&  14.521 &  0.001 &  0.618 &  0.883 &  1.038 & 250.5074 &  36.3896 & -242.0 &   1 &  99 \\
2174 &  53521 &   500&  14.756 &  0.003 &  0.620 &  0.880 &  1.039 & 250.4441 &  36.5013 & -249.6 &   1 &  99 \\
2174 &  53521 &   522&  14.977 &  0.003 &  0.589 &  0.843 &  0.980 & 250.5715 &  36.5259 & -244.9 &   1 &  99 \\
2174 &  53521 &   524&  14.495 &  0.002 &  0.620 &  0.900 &  1.036 & 250.5826 &  36.4961 & -245.9 &   1 &  99 \\
2174 &  53521 &   529&  16.009 &  0.006 &  0.533 &  0.748 &  0.875 & 250.6123 &  36.6507 & -245.6 &   1 &  \nodata \\
2174 &  53521 &   530&  14.717 &  0.003 &  0.617 &  0.876 &  1.033 & 250.5795 &  36.6176 & -246.8 &   1 &  99 \\
2174 &  53521 &   531&  14.674 &  0.004 &  0.620 &  0.886 &  1.020 & 250.6084 &  36.4513 & -243.9 &   1 &  99 \\
2174 &  53521 &   532&  15.252 &  0.002 & -0.220 & -0.361 & -0.464 & 250.5847 &  36.4509 & -245.9 &   1 &  99 \\
2174 &  53521 &   542&  14.709 &  0.003 &  0.640 &  0.890 &  1.072 & 250.4867 &  36.6975 & -248.1 &   1 &  \nodata \\
2185 &  53532 &   181&  18.574 &  0.003 &  0.282 &  0.376 &  0.353 & 250.2266 &  36.2189 & -247.5 &   1 &  \nodata \\
 2185&  53532 &   421&  18.919& 0.012&  -0.520&  -0.859&  -1.232 & 250.3562 &  36.6878& -246.4 &  2& \nodata \\
 2185&  53532 &   461&  18.172& 0.011&   0.258&   0.359&   0.333 & 250.3284 &  36.7000& -241.4 &  2& \nodata \\
2185 &  53532 &   462&  18.011 &  0.003 &  0.270 &  0.391 &  0.450 & 250.2369 &  36.7179 & -248.1 &   1 &  \nodata \\
2185 &  53532 &   466&  18.649 &  0.003 &  0.233 &  0.316 &  0.376 & 250.2879 &  36.7253 & -245.5 &   1 &  \nodata \\
 2185&  53532 &   469&  18.783& 0.012&   0.256&   0.368&   0.403 & 250.2945 &  36.6066& -246.2 &  2& \nodata \\
2185 &  53532 &   475&  18.428 &  0.001 &  0.244 &  0.338 &  0.364 & 250.2399 &  36.5889 & -249.1 &   1 &  \nodata \\
2185 &  53532 &   476&  18.883 &  0.002 &  0.267 &  0.376 &  0.396 & 250.2486 &  36.5748 & -244.4 &   1 &  \nodata \\
2185 &  53532 &   477&  19.381 &  0.002 &  0.309 &  0.415 &  0.480 & 250.2305 &  36.6108 & -244.1 &   1 &  \nodata \\
 2185&  53532 &   478&  18.732& 0.023&   0.333&   0.432&   0.452 & 250.2752 &  36.6187& -243.6 &  2& \nodata \\
2185 &  53532 &   480&  19.474 &  0.005 &  0.313 &  0.429 &  0.490 & 250.2232 &  36.6269 & -243.0 &   1 &  \nodata \\
 2185&  53532 &   481&  17.961& 0.012&   0.294&   0.422&   0.475 & 250.3259 &  36.6554& -244.6 &  2& \nodata \\
 2185&  53532 &   485&  18.251& 0.009&   0.254&   0.397&   0.376 & 250.3427 &  36.6378& -248.3 &  2& \nodata \\
 2185&  53532 &   487&  18.081& 0.009&   0.293&   0.363&   0.400 & 250.3905 &  36.5915& -242.1 &  2& \nodata \\
 2185&  53532 &   488&  19.313& 0.018&   0.294&   0.449&   0.422 & 250.3803 &  36.6619& -239.6 &  2& \nodata \\
 2185&  53532 &   489&  18.848& 0.013&   0.241&   0.360&   0.468 & 250.3864 &  36.7119& -244.6 &  2& \nodata \\
 2185&  53532 &   490&  19.105& 0.013&   0.292&   0.389&   0.377 & 250.4196 &  36.5921& -242.2 &  2& \nodata \\
 2185&  53532 &   492&  17.916& 0.011&   0.308&   0.423&   0.405 & 250.3583 &  36.6074& -240.5 &  2& \nodata \\
 2185&  53532 &   494&  18.957& 0.014&   0.310&   0.375&   0.348 & 250.4334 &  36.6197& -243.8 &  2& \nodata \\
 2185&  53532 &   495&  18.462& 0.012&   0.253&   0.357&   0.376 & 250.3132 &  36.6427& -242.4 &  2& \nodata \\
 2185&  53532 &   496&  19.600& 0.017&   0.355&   0.473&   0.395 & 250.3284 &  36.6039& -245.6 &  2& \nodata \\
2185 &  53532 &   504&  18.561 &  0.004 &  0.264 &  0.345 &  0.360 & 250.5587 &  36.6806 & -250.2 &   1 &  \nodata \\
 2185&  53532 &   507&  18.897& 0.016&   0.324&   0.402&   0.443 & 250.4724 &  36.6777& -243.3 &  2& \nodata \\
2185 &  53532 &   508&  18.818 &  0.008 &  0.286 &  0.380 &  0.433 & 250.4435 &  36.7130 & -249.1 &   1 &  \nodata \\
 2185&  53532 &   516&  19.603& 0.020&   0.321&   0.485&   0.545 & 250.4559 &  36.6134& -239.8 &  2& \nodata \\
2185 &  53532 &   545&  18.872 &  0.003 &  0.290 &  0.422 &  0.398 & 250.6272 &  36.7145 & -242.1 &   1 &  \nodata \\
2255 &  53565 &   116&  15.802 &  0.002 &  0.536 &  0.765 &  0.881 & 250.5545 &  36.2678 & -249.5 &   1 &  \nodata \\
2255 &  53565 &   118&  14.445 &  0.004 &  0.618 &  0.893 &  1.037 & 250.5464 &  36.3062 & -247.3 &   1 &  99 \\
2255 &  53565 &   144&  15.112 &  0.004 &  0.586 &  0.822 &  0.957 & 250.4085 &  36.3039 & -246.9 &   1 &  99 \\
2255 &  53565 &   147&  15.412 &  0.004 & -0.254 & -0.440 & -0.548 & 250.4520 &  36.3018 & -249.1 &   1 &  99 \\
2255 &  53565 &   148&  15.329 &  0.009 &  0.539 &  0.784 &  0.906 & 250.4910 &  36.3083 & -241.8 &   1 &   7 \\
2255 &  53565 &   152&  12.840 &  0.001 &  0.847 &  1.200 &  1.414 & 250.4111 &  36.3777 & -249.9 &   1 &  99 \\
2255 &  53565 &   153&  15.171 &  0.004 &  0.572 &  0.820 &  0.933 & 250.4041 &  36.3515 & -249.1 &   1 &  99 \\
2255 &  53565 &   156&  15.511 &  0.004 & -0.286 & -0.468 & -0.588 & 250.4906 &  36.3635 & -244.8 &   1 &  98 \\
2255 &  53565 &   157&  14.510 &  0.002 &  0.631 &  0.892 &  1.057 & 250.4289 &  36.3301 & -245.4 &   1 &  99 \\
2255 &  53565 &   160&  14.601 &  0.022 &  0.595 &  0.878 &  1.044 & 250.4894 &  36.3321 & -249.9 &   1 &  99 \\
2255 &  53565 &   167&  14.042 &  0.002 &  0.676 &  0.965 &  1.128 & 250.2349 &  36.3718 & -246.6 &   1 &  99 \\
2255 &  53565 &   172&  14.493 &  0.004 &  0.614 &  0.880 &  0.972 & 250.3615 &  36.3904 & -248.9 &   1 &  99 \\
2255 &  53565 &   174&  15.175 &  0.003 & -0.199 & -0.340 & -0.437 & 250.3136 &  36.3878 & -246.1 &   1 &  99 \\
2255 &  53565 &   175&  15.232 &  0.002 &  0.577 &  0.824 &  0.919 & 250.3261 &  36.3471 & -242.6 &   1 &  99 \\
2255 &  53565 &   178&  13.710 &  0.002 &  0.721 &  1.019 &  1.206 & 250.4017 &  36.2855 & -242.2 &   1 &  99 \\
2255 &  53565 &   426&  15.024 &  0.001 &  0.571 &  0.831 &  0.989 & 250.3146 &  36.5174 & -244.4 &   1 &  99 \\
2255 &  53565 &   428&  14.245 &  0.007 &  0.672 &  0.957 &  1.123 & 250.2309 &  36.5953 & -250.2 &   1 &  99 \\
2255 &  53565 &   431&  13.412 &  0.021 &  0.728 &  1.065 &  1.255 & 250.3326 &  36.4106 & -250.0 &   1 &  99 \\
2255 &  53565 &   433&  15.370 &  0.005 & -0.263 & -0.450 & -0.555 & 250.2911 &  36.5694 & -241.9 &   1 &  99 \\
2255 &  53565 &   436&  14.753 &  0.003 &  0.619 &  0.885 &  1.026 & 250.3078 &  36.4174 & -241.8 &   1 &  99 \\
2255 &  53565 &   440&  14.721 &  0.003 &  0.616 &  0.874 &  1.023 & 250.2608 &  36.4377 & -247.5 &   1 &  99 \\
2255 &  53565 &   464&  13.338 &  0.003 &  0.703 &  1.005 &  1.178 & 250.3784 &  36.5035 & -244.4 &   1 &  99 \\
2255 &  53565 &   465&  15.230 &  0.004 & -0.209 & -0.366 & -0.484 & 250.4422 &  36.4292 & -245.7 &   1 &  99 \\
2255 &  53565 &   468&  15.375 &  0.008 &  0.538 &  0.810 &  0.891 & 250.3910 &  36.4529 & -242.3 &   1 &  98 \\
2255 &  53565 &   472&  14.576 &  0.002 &  0.621 &  0.888 &  1.037 & 250.3964 &  36.4008 & -241.8 &   1 &  99 \\
2255 &  53565 &   474&  13.116 &  0.003 &  0.797 &  1.132 &  1.301 & 250.3681 &  36.4510 & -243.0 &   1 &  98 \\
2255 &  53565 &   475&  13.420 &  0.002 &  0.753 &  1.087 &  1.280 & 250.3134 &  36.4899 & -244.4 &   1 &  99 \\
2255 &  53565 &   477&  13.740 &  0.002 &  0.697 &  0.997 &  1.154 & 250.3787 &  36.4254 & -249.1 &   1 &  99 \\
2255 &  53565 &   482&  15.116 &  0.002 & -0.187 & -0.328 & -0.421 & 250.3339 &  36.6145 & -250.0 &   1 &  99 \\
2255 &  53565 &   486&  14.665 &  0.001 &  0.611 &  0.886 &  1.050 & 250.3160 &  36.5549 & -243.1 &   1 &  99 \\
2255 &  53565 &   487&  12.527 &  0.009 &  0.917 &  1.283 &  1.512 & 250.4381 &  36.4702 & -239.1 &   1 &  99 \\
2255 &  53565 &   490&  15.825 &  0.003 &  0.517 &  0.755 &  0.889 & 250.3626 &  36.5661 & -244.7 &   1 &  \nodata \\
2255 &  53565 &   491&  11.564 &  0.002 &  1.277 &  1.791 &  2.111 & 250.4595 &  36.4042 & -247.7 &   1 &  99 \\
2255 &  53565 &   493&  13.683 &  0.013 &  0.722 &  1.035 &  1.204 & 250.4420 &  36.4546 & -248.3 &   1 &  99 \\
2255 &  53565 &   496&  14.258 &  0.007 &  0.649 &  0.941 &  1.100 & 250.3886 &  36.5412 & -248.9 &   1 &  99 \\
2255 &  53565 &   498&  11.446 &  0.014 &  1.352 &  1.938 &  2.281 & 250.4247 &  36.4476 & -247.8 &   1 &  99 \\
2255 &  53565 &   499&  13.324 &  0.001 &  0.771 &  1.090 &  1.290 & 250.4365 &  36.3909 & -243.3 &   1 &  99 \\
2255 &  53565 &   501&  13.201 &  0.004 &  0.783 &  1.128 &  1.314 & 250.4386 &  36.5185 & -243.5 &   1 &  99 \\
2255 &  53565 &   503&  13.964 &  0.002 &  0.687 &  0.993 &  1.145 & 250.4857 &  36.5007 & -246.4 &   1 &  99 \\
2255 &  53565 &   504&  13.826 &  0.004 &  0.705 &  1.019 &  1.167 & 250.4200 &  36.5698 & -248.5 &   1 &  99 \\
2255 &  53565 &   505&  15.367 &  0.003 & -0.231 & -0.388 & -0.488 & 250.4430 &  36.5538 & -246.8 &   1 &  99 \\
2255 &  53565 &   510&  15.115 &  0.001 &  0.595 &  0.867 &  0.999 & 250.4508 &  36.5948 & -248.2 &   1 &  99 \\
2255 &  53565 &   516&  11.578 &  0.011 &  1.332 &  1.893 &  2.207 & 250.4620 &  36.4817 & -242.5 &   1 &  99 \\
2255 &  53565 &   517&  13.955 &  0.010 &  0.687 &  1.002 &  1.124 & 250.4654 &  36.4590 & -241.8 &   1 &  99 \\
2255 &  53565 &   519&  14.040 &  0.002 &  0.680 &  0.970 &  1.129 & 250.5015 &  36.4235 & -243.7 &   1 &  99 \\
2255 &  53565 &   520&  14.373 &  0.004 &  0.644 &  0.930 &  1.042 & 250.4717 &  36.4231 & -241.2 &   1 &  99 \\
2255 &  53565 &   542&  14.977 &  0.003 &  0.589 &  0.843 &  0.980 & 250.5715 &  36.5259 & -246.9 &   1 &  99 \\
2255 &  53565 &   544&  14.561 &  0.004 &  0.637 &  0.912 &  1.067 & 250.5411 &  36.4955 & -249.4 &   1 &  99 \\
2255 &  53565 &   547&  14.579 &  0.003 &  0.633 &  0.910 &  1.043 & 250.5786 &  36.5043 & -249.4 &   1 &  99 \\
2255 &  53565 &   548&  14.836 &  0.002 &  0.629 &  0.887 &  1.037 & 250.5392 &  36.5664 & -245.2 &   1 &  99 \\
2255 &  53565 &   549&  14.523 &  0.005 &  0.637 &  0.908 &  1.051 & 250.5105 &  36.5424 & -249.2 &   1 &  99 \\
2255 &  53565 &   550&  15.197 &  0.004 &  0.589 &  0.842 &  0.956 & 250.5207 &  36.5268 & -246.5 &   1 &  99 \\
2255 &  53565 &   551&  14.674 &  0.004 &  0.620 &  0.886 &  1.020 & 250.6084 &  36.4513 & -243.3 &   1 &  99 \\
2255 &  53565 &   552&  13.934 &  0.002 &  0.689 &  0.988 &  1.149 & 250.5688 &  36.4162 & -241.4 &   1 &  99 \\
2255 &  53565 &   553&  14.842 &  0.004 &  0.606 &  0.864 &  1.013 & 250.5565 &  36.4768 & -239.5 &   1 &  99 \\
2255 &  53565 &   557&  14.463 &  0.002 &  0.635 &  0.906 &  1.054 & 250.5687 &  36.4371 & -239.6 &   1 &  99 \\
2255 &  53565 &   589&  14.717 &  0.003 &  0.617 &  0.876 &  1.033 & 250.5795 &  36.6176 & -245.9 &   1 &  99 \\
2255 &  53565 &   600&  14.507 &  0.001 &  0.651 &  0.910 &  1.078 & 250.5424 &  36.6308 & -248.7 &   1 &  98 \\
2255 &  53565 &   500&  14.532& 0.005&   0.621&   0.913&   1.029 & 250.3637 &  36.5395& -240.6 &  2&  99	   \\

\enddata
     
\tablerefs{1. \citet{clem08}, 2. \citet{an08}}
     
\end{deluxetable*}

\begin{deluxetable*}{rrrrrrrrrrrrcc}
\tablewidth{0pt}
\tablecolumns{14}
\tabletypesize{\tiny}
\tablecaption{M71 members observed by SEGUE \label{m71tab}}
\tablehead{\colhead{Plate} & \colhead{MJD} & \colhead{Fiber} &
  \colhead{$r$} & \colhead{error} & \colhead{$g-r$} & \colhead{error} & \colhead{E(B--V)} &
\colhead{RA } &\colhead{Dec} &\colhead{Velocity} &\colhead{error} &\colhead{PM prob} & \colhead{ID} \\
 &  &  &
 (mag.) &(mag.) &(mag.) &(mag.) & \colhead{offset} &
\colhead{(2000)} &\colhead{ (2000)} &\colhead{(km s$^{-1}$)} &\colhead{(km s$^{-1}$)} &\colhead{(\%)}& \\
} 
\startdata

 2333  & 53682 & 163  & 11.766  & \nodata & 1.480 & \nodata & --0.01  &  298.4511  &   18.8007 &  --20.5 &  0.2& 97  & 1-45 \\
 2333  & 53682 & 167  & 14.106  & 0.002   & 0.828 &  0.004  &  0.00  &  298.4630  &   18.7697 &  --26.2 &  0.6& 85  & 1-19 \\
 2333  & 53682 & 173  & 12.477  & 0.002   & 1.273 &  0.004  &  0.00  &  298.4610  &   18.8189 &  --24.9 &  0.3& 97  & 1-53  \\
 2333  & 53682 & 185  & 12.886  & \nodata & 1.051 & \nodata & --0.02  &  298.4210  &   18.7683 &  --23.2 &  0.5& 96  & 1-95 \\ 
 2333  & 53682 & 191  & 14.203  & 0.004   & 1.069 &  0.005  &  0.04  &  298.4241  &   18.8108 &  --25.3 &  0.8& 95  & 1-59 \\
 2333  & 53682 & 224  & 13.695  & 0.004   & 1.114 &  0.008  &  0.03  &  298.3947  &   18.7727 &  --23.4 &  0.4& 95  & KC-39 \\
 2333  & 53682 & 225  & 12.407  & 0.003   & 1.347 &  0.004  &  0.05  &  298.4062  &   18.7500 &  --26.2 &  0.3& 94  & A9   \\
 2333  & 53682 & 239  & 12.072  & \nodata & 1.488 & \nodata &  0.02  &  298.4066  &   18.7914 &  --26.7 &  0.3& 90  & 1-77 \\ \\
 2338  & 53679 & 150  & 12.729  & 0.004   & 1.104 &  0.004  &  0.00  &  298.4511  &   18.8071 &  --21.6 &  0.5& 97  & 1-56 \\

\enddata
     
\end{deluxetable*}

\begin{deluxetable*}{c | rrrrrrrrrr | }
\tablewidth{0pt}
\tablecolumns{11}
\tabletypesize{\scriptsize}
\tablecaption{Variable Reddening Values (in E(B-V)) for M71 \label{redofftab} }
\tablehead{
\colhead{Dec offset} & \multicolumn{10}{c}{RA offset (arcsec)} \\
\colhead{(arcsec)} & \colhead{--225}& \colhead{--175} &\colhead{--125}& 
\colhead{--75} & \colhead{--25} & \colhead{25}& \colhead{75}& \colhead{125}&
 \colhead{175}& \colhead{225}
 }

\startdata

--225 & 0.03  & 0.02 &  0.00  & -0.01 & -0.01 &  0.00  &  0.00  &  0.00  &  0.00  &  0.03 \\
--175  & 0.04  & 0.03 &  0.03 &  0.01 & -0.01 & -0.01 &  0.00  &  0.00  &  0.00  &  0.02 \\
--125  & 0.02  & 0.03 &  0.08 &  0.02 & -0.02 & -0.02 & -0.01 &  0.00  &  0.01 &  0.03 \\
--75  & 0.03  & 0.02 &  0.02 & -0.02 & -0.02 & -0.02 & -0.02 &  0.00  &  0.02 &  0.02 \\
--25  & 0.03  & 0.03 &  0.03 & -0.02 &  0.00  &  0.00  &  0.00  &  0.00  &  0.02 &  0.01 \\
25   & 0.04  & 0.04 &  0.02 &  0.01 &  0.01 &  0.00  &  0.00  &  0.01 &  0.00  &  0.00  \\
75   & 0.04  & 0.04 &  0.03 &  0.02 &  0.02 & -0.01 & -0.01 &  0.0  &  0.0  & -0.01 \\
125   & 0.05  & 0.05 &  0.05 &  0.04 & -0.01 &  0.00  &  0.00  &  0.00  &  0.00  & -0.01 \\
175   & 0.06  & 0.05 &  0.04 &  0.02 &  0.00  &  0.00  &  0.00  &  0.00  &  0.00  &  0.00  \\
225   & 0.07  & 0.06 &  0.05 &  0.05 &  0.02 &  0.00  &  0.00  & -0.01 & -0.01 & -0.01 \\

\enddata

\end{deluxetable*}

\begin{deluxetable*}{rrrrrrrrcr}
\tablewidth{0pt}
\tablecolumns{10}
\tabletypesize{\tiny}
\tablecaption{NGC 7789 Members Observed by SEGUE \label{n7789mem}}
\tablehead{\colhead{Plate} & \colhead{MJD} & \colhead{Fiber} &
  \colhead{$V$} & \colhead{$V-I$}& $(g-r)_0$  &\colhead{RA} &\colhead{Dec} & \colhead{$E(B-V)$} &\colhead{Velocity}  \\
  \colhead{} & \colhead{} &\colhead{} 
&\colhead{(mag.)} &\colhead{(mag.)} & \colhead{(transf)}& \colhead{(2000)} & \colhead{(2000)} & \colhead{} &\colhead{(km s$^{-1}$)}  \\
} 
\startdata

2377& 53991 & 151  & 12.794  & 1.336 &  0.820  &    359.5293   &    56.6800  &  0.25&  --49.1 \\
2377& 53991 & 162  & 12.839  & 1.284 &  0.765  &    359.3542   &    56.6434  &  0.25&  --46.7 \\
2377& 53991 & 175  & 12.913  & 1.288 &  0.730  &    359.3507   &    56.6600  &  0.27&  --50.6 \\
2377& 53991 & 176  & 13.055  & 1.306 &  0.800  &    359.3866   &    56.6721  &  0.25&  --48.0 \\
2377& 53991 & 178  & 12.962  & 1.331 &  0.795  &    359.3378   &    56.5841  &  0.24&  --47.6 \\
2377& 53991 & 191  & 12.981  & 1.310 &  0.763  &    359.2386   &    56.6153  &  0.25&  --44.4 \\
2377& 53991 & 200  & 12.273  & 1.539 &  1.012  &    359.2176   &    56.5608  &  0.25&  --44.6 \\
2377& 53991 & 232  & 12.188  & 1.528 &  1.048  &    358.9569   &    56.6550  &  0.25&  --48.1 \\
2377& 53991 & 439  & 13.260  & 1.316 &  0.794  &    359.1707   &    56.6964  &  0.25&  --46.8 \\
2377& 53991 & 461  & 12.305  & 1.540 &  1.016  &    359.2311   &    56.7525  &  0.27&  --46.2 \\
2377& 53991 & 489  & 12.982  & 1.344 &  0.824  &    359.2382   &    56.6971  &  0.25&  --50.7 \\
2377& 53991 & 493  & 13.111  & 1.328 &  0.823  &    359.2648   &    56.7225  &  0.25&  --47.5 \\
2377& 53991 & 494  & 13.160  & 1.349 &  0.829  &    359.1855   &    56.7149  &  0.25&  --47.6 \\
2377& 53991 & 506  & 11.986  & 1.617 &  1.109  &    359.5007   &    56.8368  &  0.26&  --47.1 \\
2377& 53991 & 515  & 13.164  & 1.377 &  0.844  &    359.4409   &    56.8448  &  0.25&  --47.6 \\

\enddata
      
\end{deluxetable*}

\begin{deluxetable*}{rrrrrrrrrrrc}
\tablewidth{0pt}
\tablecolumns{12}
\tabletypesize{\tiny}
\tablecaption{NGC 6791 Members Observed by SEGUE \label{n6791mem}}
\tablehead{\colhead{Plate} & \colhead{MJD} & \colhead{Fiber} &
  \colhead{$r$} & \colhead{error} &\colhead{$g-r$} &\colhead{$g-i$} &\colhead{$g-z$} &
\colhead{RA } &\colhead{Dec} &\colhead{Velocity} &\colhead{PM prob} \\
 &  &  &
  \colhead{(mag.)} & \colhead{(mag.)} &\colhead{(mag.)} &\colhead{(mag.)} &\colhead{(mag.)} &\colhead{(2000)} &\colhead{ (2000)} &\colhead{(km s$^{-1}$)}  &\colhead{(\%)} \\
} 
\startdata

 2800 & 54326  & 151 & 16.943  & 0.005  &    0.927 &   1.235  &  1.439  &  290.3105 &   37.7758 &  -47.6  & 88  \\
 2800 & 54326  & 152 & 17.904  & 0.011  &    0.687 &   0.915  &  1.070  &  290.2779 &   37.8023 &  -45.8  & 96  \\
 2800 & 54326  & 154 & 15.327  & 0.006  &    1.100 &   1.495  &  1.745  &  290.2560 &   37.8014 &  -47.1  & 90  \\
 2800 & 54326  & 156 & 16.112  & 0.005  &    1.040 &   1.395  &  1.639  &  290.2894 &   37.7840 &  -43.9  & 75  \\
 2800 & 54326  & 159 & 14.061  & 0.005  &    1.118 &   1.528  &  1.762  &  290.2762 &   37.7499 &  -45.8  & 99  \\
 2800 & 54326  & 160 & 15.923  & 0.004  &    1.018 &   1.369  &  1.577  &  290.3084 &   37.7526 &  -47.2  & 85  \\
 2800 & 54326  & 161 & 17.126  & 0.007  &    0.893 &   1.195  &  1.361  &  290.2689 &   37.7212 &  -46.5  & 97  \\
 2800 & 54326  & 167 & 13.923  & 0.007  &    1.329 &  \nodata  &  2.178  &  290.2547 &   37.7037 &  -47.7  & 68  \\
 2800 & 54326  & 169 & 14.148  & 0.007  &    1.160 &   1.583  &  1.820  &  290.2448 &   37.7203 &  -45.3  & 77  \\
 2800 & 54326  & 170 & 13.545  & 0.006  &    1.440 &  \nodata  &  2.484  &  290.2191 &   37.7412 &  -47.9  & 92  \\
 2800 & 54326  & 172 & 17.284  & 0.006  &    0.736 &   0.955  &  1.086  &  290.2085 &   37.7977 &  -45.3  & 79  \\
 2800 & 54326  & 173 & 17.085  & 0.007  &    0.972 &   1.314  &  1.497  &  290.2308 &   37.7971 &  -47.9  & 81  \\
 2800 & 54326  & 174 & 14.073  & 0.005  &    1.195 &   1.625  &  1.895  &  290.2536 &   37.7594 &  -46.8  & 97  \\
 2800 & 54326  & 175 & 14.109  & 0.005  &    1.166 &   1.572  &  1.836  &  290.2536 &   37.7777 &  -49.5  & 97  \\
 2800 & 54326  & 180 & 14.095  & 0.004  &    1.126 &   1.520  &  1.767  &  290.2203 &   37.7592 &  -45.5  & 93  \\
 2800 & 54326  & 181 & 14.421  & 0.003  &    1.204 &   1.656  &  1.926  &  290.1882 &   37.7428 &  -49.6  & 98  \\
 2800 & 54326  & 182 & 17.435  & 0.007  &    0.668 &   0.893  &  0.968  &  290.1303 &   37.7752 &  -47.3  & 98  \\
 2800 & 54326  & 183 & 14.143  & 0.006  &    1.145 &   1.541  &  1.812  &  290.1889 &   37.7883 &  -52.4  & 88  \\
 2800 & 54326  & 184 & 17.368  & 0.009  &    0.657 &   0.878  &  0.979  &  290.1277 &   37.7548 &  -50.6  & 98  \\
 2800 & 54326  & 185 & 14.387  & 0.005  &    1.142 &   1.527  &  1.775  &  290.1635 &   37.7437 &  -46.8  & 98  \\
 2800 & 54326  & 187 & 17.154  & 0.007  &    0.751 &   0.998  &  1.146  &  290.1433 &   37.8011 &  -47.0  & 73  \\
 2800 & 54326  & 188 & 17.932  & 0.007  &    0.704 &   0.933  &  1.052  &  290.1961 &   37.7612 &  -45.8  & 91  \\
 2800 & 54326  & 189 & 16.560  & 0.009  &    0.992 &   1.324  &  1.518  &  290.1688 &   37.7852 &  -45.9  & 55  \\
 2800 & 54326  & 190 & 15.701  & 0.006  &    1.064 &   1.439  &  1.672  &  290.1767 &   37.7642 &  -46.3  & 88  \\
 2800 & 54326  & 194 & 17.300  & 0.011  &    0.642 &   0.852  &  0.968  &  290.1742 &   37.7060 &  -52.4  & 77  \\
 2800 & 54326  & 197 & 15.979  & 0.007  &    1.025 &   1.377  &  1.606  &  290.1808 &   37.7214 &  -46.7  & 92  \\
 2800 & 54326  & 199 & 14.094  & 0.004  &    1.287 &  \nodata<  &  2.107  &  290.1639 &   37.8013 &  -45.7  & 94  \\
 2800 & 54326  & 431 & 17.103  & 0.005  &    0.736 &   0.993  &  1.109  &  290.1247 &   37.8115 &  -40.9  & 98  \\
 2800 & 54326  & 465 & 14.972  & 0.004  &    1.139 &   1.570  &  1.844  &  290.2405 &   37.8170 &  -44.9  & 85  \\
 2800 & 54326  & 475 & 14.774  & 0.010  &    1.128 &   1.540  &  1.823  &  290.1927 &   37.8196 &  -43.2  & 99  \\
 2800 & 54326  & 479 & 17.874  & 0.008  &    0.691 &   0.947  &  1.064  &  290.1633 &   37.8346 &  -46.8  & 95  \\
 2800 & 54326  & 480 & 17.249  & 0.007  &    0.759 &   1.039  &  1.180  &  290.1579 &   37.8190 &  -44.4  & 82  \\
 2821 & 54393  & 141 & 17.951  & 0.008  &    0.721 &   0.946  &  1.066  &  290.2928 &   37.7322 &  -46.8  & 89  \\
 2821 & 54393  & 142 & 18.703  & 0.010  &    0.839 &   1.086  &  1.255  &  290.2954 &   37.7891 &  -40.9  & 96  \\
 2821 & 54393  & 145 & 17.970  & 0.008  &    0.672 &   0.911  &  0.956  &  290.3149 &   37.7871 &  -44.0  & 68  \\
 2821 & 54393  & 146 & 18.216  & 0.009  &    0.707 &   0.936  &  1.065  &  290.2860 &   37.7175 &  -43.9  & 68  \\
 2821 & 54393  & 149 & 17.586  & 0.006  &    0.674 &   0.878  &  0.966  &  290.2926 &   37.7521 &  -43.2  & 70  \\
 2821 & 54393  & 161 & 17.537  & 0.008  &    0.661 &   0.865  &  0.976  &  290.2675 &   37.7325 &  -48.9  & 83  \\
 2821 & 54393  & 165 & 18.742  & 0.012  &    0.822 &   1.119  &  1.234  &  290.1696 &   37.7074 &  -48.6  & 67  \\
 2821 & 54393  & 166 & 18.509  & 0.010  &    0.776 &   1.027  &  1.210  &  290.2165 &   37.7927 &  -43.8  & 85  \\
 2821 & 54393  & 167 & 18.392  & 0.013  &    0.772 &   1.037  &  1.095  &  290.2531 &   37.7614 &  -43.8  & 91  \\
 2821 & 54393  & 169 & 18.398  & 0.013  &    0.777 &   1.042  &  1.203  &  290.2357 &   37.7495 &  -43.6  & 94  \\
 2821 & 54393  & 172 & 18.174  & 0.011  &    0.754 &   0.991  &  1.120  &  290.2331 &   37.7795 &  -48.4  & 69  \\
 2821 & 54393  & 173 & 18.094  & 0.013  &    0.715 &   0.941  &  1.043  &  290.2340 &   37.7255 &  -46.5  & 76  \\
 2821 & 54393  & 174 & 18.263  & 0.014  &    0.736 &   0.934  &  1.075  &  290.2744 &   37.7682 &  -46.9  & 80  \\
 2821 & 54393  & 176 & 18.983  & 0.014  &    0.850 &   1.154  &  1.265  &  290.2389 &   37.7979 &  -42.3  & 63  \\
 2821 & 54393  & 177 & 18.312  & 0.011  &    0.757 &   1.000  &  1.151  &  290.2552 &   37.7811 &  -44.8  & 92  \\
 2821 & 54393  & 178 & 18.462  & 0.011  &    0.787 &   1.051  &  1.215  &  290.2708 &   37.7936 &  -47.7  & 84  \\
 2821 & 54393  & 179 & 17.434  & 0.013  &    0.698 &   0.925  &  1.037  &  290.2332 &   37.6950 &  -46.3  & 61  \\
 2821 & 54393  & 182 & 18.009  & 0.008  &    0.738 &   0.964  &  1.044  &  290.1917 &   37.7502 &  -44.9  & 98  \\
 2821 & 54393  & 183 & 17.738  & 0.009  &    0.668 &   0.870  &  0.983  &  290.1618 &   37.7224 &  -48.2  & 68  \\
 2821 & 54393  & 187 & 17.646  & 0.009  &    0.646 &   0.868  &  0.972  &  290.1851 &   37.7333 &  -49.2  & 72  \\
 2821 & 54393  & 188 & 18.770  & 0.010  &    0.878 &   1.157  &  1.295  &  290.1615 &   37.7461 &  -42.9  & 72  \\
 2821 & 54393  & 190 & 19.206  & 0.018  &    0.969 &   1.265  &  1.434  &  290.2118 &   37.7134 &  -43.0  & 85  \\
 2821 & 54393  & 191 & 18.251  & 0.011  &    0.732 &   0.980  &  1.100  &  290.1620 &   37.7770 &  -50.4  & 95  \\
 2821 & 54393  & 193 & 18.495  & 0.009  &    0.814 &   1.069  &  1.168  &  290.1486 &   37.7595 &  -44.2  & 97  \\
 2821 & 54393  & 194 & 18.047  & 0.009  &    0.737 &   0.971  &  1.087  &  290.1838 &   37.7774 &  -45.6  & 95  \\
 2821 & 54393  & 195 & 18.852  & 0.012  &    0.842 &   1.122  &  1.293  &  290.2029 &   37.7670 &  -44.5  & 89  \\
 2821 & 54393  & 196 & 18.058  & 0.012  &    0.708 &   0.939  &  1.101  &  290.2135 &   37.7412 &  -48.2  & 97  \\
 2821 & 54393  & 197 & 17.757  & 0.006  &    0.664 &   0.888  &  0.988  &  290.1443 &   37.7864 &  -47.3  & 94  \\
 2821 & 54393  & 198 & 18.678  & 0.012  &    0.819 &   1.105  &  1.209  &  290.1754 &   37.7626 &  -48.5  & 68  \\
 2821 & 54393  & 199 & 18.890  & 0.014  &    0.867 &   1.150  &  1.331  &  290.1903 &   37.7149 &  -48.0  & 72  \\
 2821 & 54393  & 200 & 18.260  & 0.011  &    0.751 &   0.974  &  1.090  &  290.1816 &   37.7957 &  -47.5  & 56  \\
 2821 & 54393  & 232 & 19.101  & 0.016  &    0.954 &   1.342  &  1.515  &  290.1246 &   37.7310 &  -46.9  & 87  \\
 2821 & 54393  & 235 & 17.550  & 0.009  &    0.689 &   0.900  &  0.991  &  290.1257 &   37.7643 &  -42.1  & 95  \\
 2821 & 54393  & 436 & 18.307  & 0.008  &    0.774 &   1.082  &  1.253  &  290.1259 &   37.8133 &  -44.3  & 91  \\
 2821 & 54393  & 439 & 17.568  & 0.006  &    0.688 &   0.926  &  1.027  &  290.1193 &   37.7980 &  -44.1  & 98  \\

\enddata

\end{deluxetable*}

\begin{deluxetable*}{rrrrrrrrrrrc}
\tablewidth{0pt}
\tablecolumns{12}
\tabletypesize{\tiny}
\tablecaption{Berkeley 29 members observed by SEGUE \label{be29tab}}
\tablehead{\colhead{Plate} & \colhead{MJD} & \colhead{Fiber} &
  \colhead{$g$} & \colhead{error} &\colhead{$g-r$} &\colhead{error} &\colhead{RA } &\colhead{Dec} &\colhead{Velocity} &\colhead{error} &\colhead{other ID \tablenotemark{a}} \\
\colhead{} & \colhead{} &\colhead{}  &
  \colhead{(mag.)} & \colhead{(mag.)} &\colhead{(mag.)} &\colhead{(mag.)} &\colhead{(2000)} &\colhead{(2000)} &\colhead{ (km s$^{-1}$))} &\colhead{(km s$^{-1}$)}  &\colhead{} \\
} 
\startdata
3335 & 54922 & 113 &  19.110 &  0.015 &  0.599  &   103.3311  &  16.8730 &  0.017 &  26.57  & 2.22 &  \\
3335 & 54922 & 195 &  16.799 &  0.007 &  0.761  &   103.1608  &  16.8460 &  0.011 &  34.56  & 1.04 &  \\
3335 & 54922 & 462 &  18.926 &  0.011 &  0.658  &   103.2417  &  16.9463 &  0.018 &  39.85  & 1.85 &  \\
3335 & 54922 & 474 &  19.288 &  0.014 &  0.597  &   103.1888  &  16.8793 &  0.021 &  18.02  & 2.80 &  \\
3335 & 54922 & 481 &  17.039 &  0.006 &  0.746  &   103.2836  &  16.9279 &  0.008 &  17.01  & 0.86 &  S398,C801,F948 \\
3335 & 54922 & 495 &  17.118 &  0.007 &  0.779  &   103.3010  &  16.9836 &  0.010 &  17.81  & 0.95 &  F1437 \\
3335 & 54922 & 496 &  17.074 &  0.006 &  0.776  &   103.2311  &  16.9610 &  0.009 &  18.13  & 0.89 &  S602 \\
3335 & 54922 & 497 &  17.120 &  0.008 &  0.778  &   103.2566  &  16.9392 &  0.010 &  18.22  & 0.85 &  S159 \\
3335 & 54922 & 498 &  18.165 &  0.007 &  0.701  &   103.2626  &  16.9245 &  0.011 &  21.69  & 1.34 &  \\
3335 & 54922 & 508 &  18.076 &  0.011 &  0.718  &   103.2837  &  16.9631 &  0.015 &  16.68  & 1.53 &  \\

\enddata

\tablenotetext{a}{Identifications are from S: \citet{ses08}, C: \citet{car04} and F: \citet{frinchaboy}.}

\end{deluxetable*}

\end{document}